\documentclass[%
reprint,
amsmath,amssymb,
aps,
pre,
]{revtex4-1}
\pdfoutput=1
\usepackage{amsmath,amsthm,amssymb}
\usepackage{graphicx}% Include figure files
\usepackage{dcolumn}% Align table columns ondecimal point
\usepackage[colorlinks=true]{hyperref}
\usepackage{bm}% bold math
\usepackage[export]{adjustbox}

\usepackage[colorinlistoftodos, textsize=small]{todonotes}

\usepackage{xcolor}
\usepackage{stackengine}
\setstackgap{L}{.5\baselineskip}

\usepackage{ifthen}

\newcommand{\lc}{\left\{}
\newcommand{\rc}{\right\}}
\newcommand{\lb}{\left(}
\newcommand{\rb}{\right)}
\newcommand{\ls}{\left[}
\newcommand{\rs}{\right]}
\newcommand{\abs}[1]{\left\vert #1\right\vert}

\newcommand{\di}{{\rm d}}
\newtheorem{theor}{{\bf Theorem}}

\begin{document}

\preprint{APS/123-QED}

\title{Transfer entropy computation using the Perron-Frobenius operator}% Force line breaks with \\

\author{David Diego}
\email{david.castro@uib.no}
\email{diegocastro.david@gmail.com}

\author{Kristian Agas{\o}ster Haaga}%
\homepage{https://www.earthsystemevolution.com}
\homepage{https://github.com/kahaaga/}
%\affiliation{%

\author{Bjarte Hannisdal}
\affiliation{
Department of Earth Science, University of Bergen\\
PO Box 7803, NO-5020 Bergen, Norway% with \\
}%

\date{\today}% It is always \today, today,

\begin{abstract}
We propose a method for computing the transfer entropy between time series using Ulam's approximation of the Perron-Frobenius (transfer) operator associated with the map generating the dynamics. Our method differs from standard transfer entropy estimators in that the invariant measure is estimated not directly from the data points but from the invariant distribution of the transfer operator approximated from the data points. For sparse time series and low embedding dimension, the transfer operator is approximated using a triangulation of the attractor, whereas for data-rich time series or higher embedding dimension we use a faster grid approach.   
We compare the performance of our methods with existing estimators such as the $k$ nearest neighbors method and kernel density estimation method, using coupled instances of well known chaotic systems: coupled logistic maps and a coupled R\"ossler-Lorenz system. We find that our estimators are robust against moderate levels of noise. For sparse time series with less than a hundred observations and low embedding dimension, our triangulation estimator shows improved ability to detect coupling directionality, relative to standard transfer entropy estimators.

% \begin{description}
% Secondary publications and information retrieval purposes.
% \item[PACS numbers]
% May be entered using the \verb+\pacs{#1}+ command.
% You may use the \texttt{description} environment to structure your abstract;
% use the optional argument of the \verb+\item+ command to give the category of each item. 
% \end{description}

\end{abstract}


\maketitle

%\tableofcontents

\section{\label{sec:intro}Introduction}
Time series analysis is used to study the dynamics of complex systems across many disciplines, including macro-scale activity of the brain \cite{Breakspear2017,*Bressler2011,*kreuz2011time,*lehnertz1998can,*elger1998seizure} and interactions in the global climate system \cite{Runge2015,*paluvs2011discerning,*ghil2002advanced,*schlesinger1994oscillation}. A long-standing problem in time series analysis is the detection of causal connections between different components of a system from observed time series. Several theoretical frameworks have been proposed to address this problem \cite{Kantz2003}, including information-theoretic approaches \cite{Hlavackova-Schindler2007,Amblard2013}.
% among them, two widely used methods are: information transfer entropy (TE) \cite{Schreiber2000} and Convergent Cross Mapping (CCM) \cite{Sugihara2012}. The latter method makes use of delay reconstructions of attractors (after the celebrated Takens's theorem \cite{Takens1981}) and estimates the degree of confidence in assuring that two given time series belong to the same underlying dynamics by probing to what extent the delay reconstruction out of one time series can be used to reconstruct the other time series. 
A popular information-theoretic method is the so-called transfer entropy (TE) \cite{Schreiber2000}, or conditional mutual information \cite{Palus2001}, which quantifies whether knowledge of changes in one variable reduces uncertainty about changes in another variable. 
For deterministic systems, the concept of information entropy relies on the existence of invariant densities associated to attractors \cite{Eckmann1985}.  Standard methods for computing mutual information (and from it, TE), estimate the invariant distribution directly from the embedding of the data. For instance, the  k nearest neighbors method (kNN) \cite{Kraskov2004}, which uses counting of nearest neighbors, or approaches using visitation frequency, either directly \cite{Schreiber2000} or through kernel density estimation (KDE) \cite{steuer2002mutual}. Other TE estimators use the concept of permutation entropy \cite{staniek2008symbolic,*dickten2014identifying,*bandt2002permutation}.
In our work, we propose to compute TE (or any standard information theoretic measure \cite{Cover2006}) between time series, based on a numerical approximation of the Perron-Frobenius (transfer) operator of the underlying dynamics. The transfer operator \cite{Lasota1994,*Berman1979,*Beck1993} dictates how densities in phase space are transformed under the action of maps and its approximation has been used to identify long-term emergent behavior in dynamical systems. Applications include computation of the stretching rate of chaotic maps, and identification of attracting regions in ocean circulation  \cite{Froyland2012,*Miron2017,*froyland2015studying,*froyland2014well,*ser2017lagrangian,*maes2018surface,*mcadam2018surface,Froyland1997}.  
The transfer operator enables estimation of invariant distributions from which information entropies may be computed \cite{froyland1999using,bollt2012synchronization,liang2005information}.
%In addition, the spirit of our work is to show the applicability of this approach to sparse time series corresponding to systems that are not strongly coupled, and thus out of synchronization. Which has a number of useful extensions.
%in addition to computing TE. 
Transfer operator approximation may also be used to generate dynamically informed surrogates for null-hypothesis testing, and to interpolate and forecast time series. These possibilities are explored in a forthcoming study \footnote{K.A. Haaga, D. Diego and B. Hannisdal. doi:10.17605/OSF.IO/M57EX}.

%Here, we use the approximation of the transfer operator to develop a TE estimator. 

In the following, we describe our approach starting with the general notion of TE, and a brief overview of entropies in the context of deterministic systems (for a more in-depth review, see \cite{Kantz2003,Eckmann1985}). 

\section{Transfer entropy}
 
Suppose that for two variables, $X_1$ and $X_2$, one is given the probability density $P(X_1=x_1,X_2=x_2)$ with support $S$. From this density one may compute their mutual information \cite{Cover2006,Shannon1948} as 

{\small \begin{equation}
I(X_1,X_2) = \int_S P(x_1,x_2)\,{\rm log}\frac{P\lb x_1,x_2\rb}{P\lb x_1\rb P\lb x_2\rb}\,.
\end{equation}}

For the case of three variables, $X_1,X_2$ and $X_3$, a related quantity is the TE
{\small \begin{equation}\label{te}
TE_{X_1,X_2|X_3} = \int_S P(x_1,x_2,x_3)\,{\rm log}\frac{P\lb x_1|x_2,x_3\rb}{P\lb x_1|x_3\rb}\,.
\end{equation}}
One can easily check the identity  
{\small \begin{equation}
TE_{X_1,X_2|X_3} = I(X_1,X_{23}) - I(X_1,X_2)\label{te_mi}\,,
\end{equation}}
\noindent with $X_{23}=(X_2,X_3)$. $TE_{X_1,X_2|X_3}$ thus quantifies the amount of information shared between $X_1$ and $(X_2,X_3)$ beyond the information already shared between $X_1$ and $X_2$. The TE was originally introduced in the context of time series analysis \cite{Schreiber2000} as a way of estimating the information transfer from one time series to another. Given two time series $X$ and $Y$, the TE  measures how much information is lost by assuming that the the variables $X$ and $Y$ are independent, i.e., by assuming that $P(x(t+\tau)|x(t),y(t))=P(x(t+\tau)|x(t))$. There is, however, no absolute scale associated with information entropy \cite{Cover2006,Shannon1948}, hence the TE only determines whether the information transfer from $Y$ to $X$ is greater than in the opposite direction.

\section{Measure theoretic entropies for deterministic systems}

In the context of dynamical systems, mutual information (or any information theoretic measure) between variables may be computed from the density distribution of invariant measures associated with the attractor of the dynamical system. In the following, we briefly review these concepts. 

\subsection{Attracting sets and attractors}

Assume the dynamics is generated by a diffeomorphism \footnote{The same would be true for a dynamical system generated by a vector field.} $\psi:\mathbb R^n\to \mathbb R^n$. 
A set $\mathbb A\subset \mathbb R^n$ is said to be an attracting set for $\psi$ if the following conditions are met \cite{Ruelle1981}: 

\begin{enumerate}
\item There is an open set  $U\supset\mathbb A$ and a natural number $N$ such that for any open set $V\supset\mathbb A$, $\psi^m(U)\subset V$ for all $m\geq N$.  
\item $\psi(\mathbb A)\subset \mathbb A$
\end{enumerate} 

\noindent Here $\psi^m$ denotes the $m$-fold iterate of $\psi$. The open set $U$ is called a fundamental neighborhood of $\mathbb A$. This definition implies \cite{Ruelle1981} that $\mathbb A=\cap_{m\geq 1} \psi^m(U)$ and $\psi(\mathbb A)=\mathbb A$. Moreover, if there is an open set $U\subset R^n$ such that for all $m$ big enough $\psi^m(U)$ has compact adherence contained in $U$, then $\mathbb A=\cap_{r\geq 1} \psi^r(U)$ is a compact attracting set with fundamental neighborhood $U$. 
Because the open set $V$ can be arbitrarily small around $\mathbb A$, all the trajectories entering $U$ asymptotically approach $\mathbb A$. In addition, $B=\cup_{m\geq 0} \psi^{-m} (U)$, where $\psi^{-m}$ denotes the pre-image of the $m$-fold iterate of $\psi$, is such that for any $p\in B$ there is $m$ with $\psi^m(p)\in U$. Thus, the corresponding orbit approaches $\mathbb A$ asymptotically. $B$ is called the basin of attraction of $\mathbb A$ and if $B=\mathbb R^n$, $\mathbb A$ is called a global attracting set. An attractor is, however, a somewhat more restrictive concept than an attracting set. Roughly speaking, an attractor is what is left of an attracting set after removing the wandering points (see \cite{Ruelle1981} for a precise definition).
Trivial examples of attractors are asymptotically stable fixed points while less trivial ones are stable limit cycles or quasi periodic limiting orbits %
\footnote{A typical example of a quasi periodic orbit is given by the map on the torus defined as $T_t(x,y) = (x+t,y+r\,t)$ [each component taken modulo 1] for $r>0$ non rational and 
$0\leq x,y<1$. It can be shown that such a trajectory returns arbitrarily close and infinitely often to itself, but it never closes up exactly.}. More complicated (`strange') attractors contain unstable orbits (i.e. sensitivity to initial conditions) and usually fractal geometries. Most dynamical systems associated with natural processes, even simple processes involving very few variables, give rise to highly complex dynamics in the form of strange attractors \cite{Lorenz1963,May1976,ruelle1971nature}. %ruelle1971nature

\subsection{Invariant measures and ergodicity}

The trajectory of a typical orbit of a dynamical system having an attractor generates a distribution of points in the phase space with a certain density which seems to be intrinsic to the system. 
Different portions of the attractor are visited by the orbit with different frequency, and this frequency of visitations naturally defines a density on the attractor, clearly invariant under the dynamics. The notion of invariance leads to the notion of {\it ergodicity}. 
Intuitively, a dynamical system is said to be ergodic if a generic trajectory fills in the attractor (according to the above notion of invariant density).  
A crucial result pertaining to ergodic systems is the celebrated Birkhoff's ergodic theorem:\\
\\
{\it Given a space $M$ and a map $h:M\to\ M$, let $\mu$ be a measure on $M$ such that $\mu$ is invariant under $h$. Then for any $\phi:M\to\mathbb R$, measurable, it holds that
\begin{equation}\label{ergodictheorem}
\lim_{n\to\infty} \frac{1}{n}\sum_{k=0}^{n-1} \phi\lb h^k(x)\rb=\int_M\phi\,\di\mu\,,
\end{equation}
for $\mu$-almost every $x\in M$.}\\
\\
\noindent Applied to the characteristic function \footnote{The characteristic function of a set $A$ returns $1$ if $x$ belongs to $A$ and $0$, otherwise.} of any measurable set, $K$ , the theorem implies that the measure of $K$, $\mu(K)$, equals the frequency of visits to $K$, in concordance with the above notion of invariant measure. In appendix \ref{app:ergodic}, we show the equality between time and spatial averages obtained from the estimates of invariant measures considered in this work.

Assuming that a density, $\delta$, of $\mu$ is well defined for $\mu$-almost all points \footnote{If $\mu$ assigns positive measure to sets when, and only when, the sets have positive volume (and then $\mu$ is said to be compatible with Lebesgue), a unique density $\delta$ of $\mu$ is guaranteed to exist by the Radon-Nikodym theorem \cite{Cohn2013}. In the case of dynamical systems, axiom-$A$ maps are known to posses a unique invariant measure compatible with Lebesgue \cite{Bowen1975,*Bowen1975a,*Ruelle1976}.} 
($\mu(K) = \int_K \delta\, \di m$, with $\di m$ the Euclidean (Riemannian) volume element induced on the attractor), any standard information theoretic entropy \cite{Cover2006} can be computed from $\delta$.

Axiom-$A$ systems are known to possess an (unique) invariant measure of physical significance \cite{Bowen1975,*Bowen1975a,*Ruelle1976} (compatible with the volume measure on the attractor). The property of being axiom-$A$ refers to the existence of a continuous and invariant splitting of the tangent space, at each point of the attractor, into stable and unstable directions \cite{smale1967differentiable}. This property is difficult (if not impossible) to check from an observed times series. However, a system with a sufficiently large number of degrees of freedom, and in a stationary state,
can be regarded, for the purpose of computing macroscopic properties, as
a smooth dynamical system with a transitive axiom-$A$ global attractor \cite{gallavotti1996chaotic,*gallavotti1995dynamical}.

\subsection{Estimation of the transfer operator and invariant measures}

Let $\psi:\mathbb R^n\to \mathbb R^n$ be differentiably invertible and $\mathbb A\subset \mathbb R^n$ a compact attractor with $m$ denoting the induced volume measure on $\mathbb A$. Suppose that $\mu$ is a measure compatible with $m$,  with support contained in $\mathbb A$ and having density $\delta$ with respect to $m$.
The map $\psi$ acts on the measure as $(\psi_*\mu)(K) := \mu(\psi^{-1}(K))$, for any measurable set $K\subset \mathbb A$. Thus its density is modified as   
$(\psi_* \delta)(x) := |\di_x \psi^{-1}|\cdot \delta\circ \psi^{-1} (x)$, $|\cdot|$ denoting the absolute value of the determinant. The linear map between functions 

\begin{equation}
\mathcal P(f) (x):= |\di_x \psi^{-1}|\cdot f\circ \psi^{-1} (x)\,, \label{PF}
\end{equation}

\noindent is known as the Perron-Frobenius (or transfer) operator associated to the map $\psi$. If $\mu$ is a $\psi$-invariant measure, then
{\small\begin{equation}
\delta(x) = |\di_x \psi^{-1}|\cdot \delta\circ \psi^{-1} (x)\,,
\end{equation}}
and thus $\psi$-invariant densities correspond to fixed points of $\mathcal P$. There is a rich literature on the approximation of the transfer operator and the estimation of invariant measures \cite{Froyland1997,Froyland2012}. 
Ulam's method \cite{Ulam1964} approximates the transfer operator by a row-stochastic Markov matrix acting on distributions defined over a given partition of $\mathbb A$. More specifically, let $\left\{B_1,\cdots,B_N\right\}$, be a partition of $\mathbb A$ into measurable sets and for each $1\leq a\leq N$, let $\chi_a:\mathbb R^n\to \mathbb R$ be defined as $\chi_a(x) = 1$ if $x\in B_a$ and $0$, otherwise. Any (measurable) function $\rho:\mathbb R^n\to \mathbb R_+$ can be approximated as constant on each partition element, that is 

{\small
\begin{equation}
    \rho = \sum_{i=1}^N \frac{\int_{B_i} {\rm d} m\, \rho}{m(B_i)}\, \chi_i\,,
\end{equation}
}

where $m(B_i)$ denotes the volume of the partition element $B_i$ and ${\rm d} m$ is the (Lebesgue) volume element. From this piecewise constant approximation and from equation (\ref{PF}) it follows that

{\small
\begin{align}
    \mathcal P(\chi_i) &= \sum_{j=1}^N\frac{1}{{m(B_j)}}\int_{B_j} {\rm d} m\, |\di_x \psi^{-1}|\,\chi_i\circ \psi^{-1}\,\,\chi_j\nonumber\\
    &=\sum_{j=1}^N\frac{\int_{\psi^{-1}(B_j)} {\rm d} m\, \chi_i}{{m(B_j)}}\,\chi_j\nonumber\\
    & = \sum_{j=1}^N\frac{m(B_i\cap \psi^{-1}(B_j)) }{{m(B_j)}}\,\chi_j\,. 
\end{align}
}

The linearity of $\mathcal P$ implies

{\small
\begin{equation}
    \tilde \rho : = \mathcal P(\rho) = \sum_{i,j} \frac{1}{m(B_j)}\rho_i \frac{m(B_i\cap \psi^{-1}(B_j)) }{{m(B_i)}}\,\chi_j\,, 
\end{equation}
}

where $\rho_i : = \int_{B_i} {\rm d} m \, \rho$ is the measure of $B_i$ according to the density $\rho$. Taking $\tilde\rho_i : = \int_{B_i} {\rm d} m\,\tilde\rho$ as the updated measure of $B_i$, one finds

{\small
\begin{equation}
    \tilde \rho_j = \sum_{i=1}^N \rho_i \frac{m(B_i\cap \psi^{-1}(B_j)) }{{m(B_i)}} \,. 
\end{equation}
}

The row stochastic matrix with entries 
{\small
\begin{equation}
    P^{(N)}_{ij} =  \frac{m(B_i\cap \psi^{-1}(B_j)) }{{m(B_i)}}\,, \label{transMatrix} 
\end{equation}
}
constitutes the Ulam's approximation to the transfer operator and it approximates how distributions defined over a given partition do change under the map generating the dynamics. Accordingly, its left invariant distribution ($\rho^{(N)}\cdot P^{(N)} = \rho^{(N)}$) corresponds to the approximation of the invariant density of the system subject to the partition.
From the left invariant distribution of $P^{(N)}$, a measure on $\mathbb A$ can be defined as

{\small
\begin{equation}
    \mu_N(E) = \sum_{i=1}^N \rho^{(N)}_i \frac{m(E\cap B_i)}{m(B_i)}\,.\label{measure}
\end{equation}
}

In \cite{Froyland1997} it is shown that using a piecewise linear approximation of the map $\psi$, the above sequence of measures, $\lc \mu_N\rc$, approaches a $\psi$-invariant measure as the partition gets infinitely refined (the maximum size of the sets in the partition approaches $0$ as $N\to\infty$). 
In the following sections we give a detailed description of the implementation of these approximations.

\subsection{Computation of TE}
\label{sec:te_comp}

Suppose $X$ and $Y$ are time series of two variables of a dynamical system with attractor $\mathbb A$. The attractor may be reconstructed using a generalized delay embedding from both time series as $(x^{(l)}(t) , y^{(k)}(t))$ \cite{takens1981detecting,*Sauer1991,*Deyle2011}. Where $x^{(l)}(t) = (x(t),\cdots,x(t-(l-1)\tau))$ and $y^{(k)}(t) = (y(t),\cdots,y(t-(k-1)\tau))$, for appropriate delay $\tau$ and embedding dimension $l+k$. Denote the resulting embedded attractor by $\tilde{\mathbb A}$. 
The transfer operator and the invariant measure 
may be approximated using equations (\ref{transMatrix}) and (\ref{measure}) in the $(x^{(l)} , y^{(k)})$ embedding space. Suppose $\tilde\mu$ is the invariant measure on $\tilde{\mathbb A}$ and $P(x^{(l)},y^{(k)})$ is the corresponding density of $\tilde\mu$. The TE from $Y\to X$, is then

{\small \begin{align}
&TE_{Y\to X} = \nonumber\\
& \int_{\tilde{\mathbb A}} \di m\,P\lb x^{(l)},y^{(k)}\rb\,{\rm log}\frac{P\lb x^{(j)}|x^{(l-j)},y^{(k)}\rb}{P\lb x^{(j)}|x^{(l-j)}\rb}\,.\label{TE}
\end{align}}

\noindent Notice that this procedure can be easily extended to compute any of the standard information theoretic measures by using the appropriate embedding. For instance, for the conditional TE, $TE_{Y\to X|Z}$, one may use a generalized embedding of the form $\lb x^{(l)}(t),y^{(m)}(t),z^{(k)}(t)\rb$ \footnote{With mild modifications in the derivation of the main result of chapter 4 in \cite{Froyland1997}, it can be shown that the estimate of the invariant measure is independent of the embedding.}. %Appendix \ref{app:embedding_inv} discusses the conditions under which the estimate of the invariant measure is independent of the embedding. 

During the revision of this manuscript we were made aware of the work by Bollt \cite{bollt2012synchronization}, who proposed to use Ulam's approximation to the transfer operator to estimate the transfer entropy between coupled systems in order to identify synchronization. In his work, Bollt interprets the transfer matrix, that constitutes the Ulam's approximation, as a conditional probability between states in the phase space and computes the TE using Bayes's rule. In our work, we follow a different strategy: we use a generalized embedding to approximate the transfer operator and the invariant distribution of this transfer operator is then interpreted as a joint probability on the phase space, from which TE is computed. Our method also differs from Bollt's in the use of a triangulation estimator for sparse time series, as described in the following section.

\subsection{Numerical implementation}

Suppose that $X = \lc x_1,\cdots,x_N\rc$ and $Y = \lc y_1,\cdots,y_N\rc$ are time series of two variables of some dynamical system generated by the map $\psi$, and that $TE_{Y\to X}$ is to be computed.
The collection of points $E=\lc (x^{(j)}_{n+k},x^{(l)}_n,y^{(r)}_n)\rc_{n=1}^N$, for $x^{(l)}_n = (x_n,\cdots,x_{n-l+1})$ (analogously for $y^{(r)}_n$) is a reconstruction of the attractor for suitable time delay $k$ and embedding dimension $j+l+r$ \cite{takens1981detecting,*Sauer1991,*Deyle2011}. 
Several methods for estimating both parameters can be found in the literature \cite{Kantz2003,Fraser1986,*Liebert1989,*Kim1999}.  Suppose for the moment that the transfer operator has already been approximated using equation (\ref{transMatrix}) and an estimate for an invariant measure for $E$, $\mu$, has been obtained from equation (\ref{measure}). For convenience we relabel the axes corresponding to $x^{(j)}_{n+k},x^{(l)}_n,y^{(r)}_n$ as $1,2,3$, respectively.
We use a regular grid into (hyper) rectangular bins, say $\lc C_i\rc_{i=1,\cdots,J}$, and uniquely decompose each bin index, $i$, into the triplet $(i_1,i_2,i_3)$ (appendix \ref{app:binning} for details). The integral expression for the TE, equation (\ref{te}), can then be approximated as 
{\small\begin{align}
TE_{Y\to X} & \simeq \sum_{i_1,i_2,i_3} m(C_i) P(i_1,i_2,i_3) \,\log \frac{P(i_1|i_2,i_3)}{P(i_1|i_2)}\,,  
\end{align}}
where $m(C_i)$ is the Euclidean volume of the $i$-th bin and $P(i_1,i_2,i_3)=\dfrac{\mu(C_i)}{m(C_i)}$, i.e. the density of $\mu$ over the bin $C_i$. Defining $\mu_{i_1i_2i_3} = \mu(C_i)$, one easily checks that
{\small\begin{align}
&\sum_{i_1,i_2,i_3} m(C_i) P(i_1,i_2,i_3) \,\log \frac{P(i_1|i_2,i_3)}{P(i_1|i_2)}  \nonumber\\
&= - H(i_1,i_2,i_3) - H(i_2) + H(i_1,i_2) + H(i_2,i_3)\,,\label{discrete_te}
\end{align}}
where $H(i_1,i_2,i_3)$ denotes the Shannon entropy of the distribution $\mu_{i_1i_2i_3}$, $H(i_1,i_2)$ is the entropy of the marginal distribution 
$\sum_k \mu_{i_1 i_2 k}$, $H(i_2)$ corresponds to the Shannon entropy of the marginal distribution 
$\sum_{l,k} \mu_{l i_2 k}$ and so on. \\
\\
To estimate the transfer operator and the invariant measure, we propose two different approaches depending on the length of the time series (the number of observations). Denote the set of points in the reconstructed attractor $E$ by $\lc p_{n}\rc_{n=1}^N$.

\subsubsection{Grid estimator.}

If the time series $X$ and $Y$ contain a sufficient number of observations \footnote{This number is likely system dependent.}, the transition matrix in equation (\ref{transMatrix}) can be approximated by a coarse-grained estimation \cite{dellnitz2001algorithms} as

\begin{equation}
P_{ij} \simeq \frac{\sharp\lc p_{n}\,|\, \psi(p_{n})\in C_j\,\cap\,p_n\in C_i\rc}{\sharp\lc p_m\,|\,p_m\in C_i\rc}\,,\label{grid_to}
\end{equation}   
 with $\sharp$ denoting the cardinal. In this case the left invariant distribution $\rho$ from equation (\ref{measure}) coincides with the measure of the bins, that is: $\rho_i = \mu(C_i) = \mu_{i_1i_2i_3}$. We clarify that the name grid estimator refers to the fact that the transfer operator is approximated using a partition into rectangular bins, as opposed to using a partition consisting of simplices, which we consider in the next section. The word grid does not imply that we use a visitation frequency estimator. For time series with a sufficient number of observations, however, the visitation frequency and the grid estimators converge to the same invariant distribution (appendix \ref{app:vf}). The motivation for obtaining an estimate of the transfer operator is that it provides an approximation to the underlying map that has applications beyond the computation of TE. For the purpose of this study, and as we show in the following sections, computing TE from the transfer operator is advantageous for sparse and noisy time series.

\subsubsection{Triangulation estimator.}\label{sec:triang}

For time series with fewer observations, the estimation of the transfer operator using equation (\ref{grid_to}) might become inaccurate. Actually, the transfer matrix obtained with the grid method for time series with few observations might fail to be Markovian. This is because the bin containing the last point in the embedding might not contain any other point. In that case such a bin is a sink of information. We thus adopt the method developed in \cite{Froyland1997}, by which the reconstructed attractor $E$ is triangulated into simplices \footnote{Built-in routines for Delaunay triangulation in arbitrary dimensions can be found in standard numerical softwares, for instance the {\it Qhull} library.} (the vertices of each simplex being points from the embedding $E$). Suppose $\lc S_1,\cdots,S_N\rc$ is such a triangulation. The map $\psi$ is then approximated by a linear map, $\tilde\psi$, on each simplex such that if $\lc p_{a_0} ,\cdots,p_{a_d}\rc$ are the vertices of the simplex $S_a$, then $\lc p_{a_0+1} ,\cdots,p_{a_d+1}\rc$ are the vertices of its image under the map, $\tilde\psi(S_a)$. The transfer matrix is obtained as

\begin{equation}
P_{ab}=\frac{m\lb S_b\cap \tilde\psi(S_a)\rb}{m(\tilde\psi(S_a))}\,,\label{MarkovMatrix}
\end{equation}

and according to equation (\ref{measure}), the measure of the simplex $a$ is $\rho_a$. To compute the simplex intersection volume we follow a direct approach outlined in appendix \ref{app:PolytopeVol}, although several methods for polytope volume computation can be found in the literature \cite{Bueler2000}.  
Once we have obtained the invariant distribution over the simplices, finding $\mu(C_i)$ exactly is computationally rather demanding. Instead, we estimate the measure of each bin by evenly sampling each simplex of the triangulation with $M_s$ points, and then assuming that each sampling point carries a fraction $1/M_s$ of the measure of the simplex they belong to. Thus, if the set of sampling points belonging to the bin $C_i$ is formed by $N_a$ points from simplex $a$, for $a=1,\cdots,N$ (possibly with some $N_a$ being zero), its measure is estimated as  
\begin{equation}
\mu(C_i) \simeq \sum_{a=1}^N \frac{N_a}{M_s}\rho_a\,.
\end{equation}

Using this sampling enables a virtually unlimited number of points to estimate the density of the measure. Notice that we do not introduce any bias by doing so (provided the sampling is even over each simplex) because the sampling points do not contain any  information beyond that encoded in $\rho$. 
 
Using embedding dimension 3 and time series with a few hundred observations, the number of simplices with positive measure out of the triangulation is on the order of hundreds (appendix \ref{app:timing}). For the examples we study here, TE becomes independent of the (total) number of sampling points beyond $\sim 5000$ \footnote{We expect this number to be dependent on the system and the embedding dimension.}. Therefore, $M_s$ will be on the order of tens. 
We subsample the simplices using a shape preserving simplex splitting routine developed in \cite{Edelsbrunner2000}. This algebraic procedure uses an edge wise splitting factor $r$ and splits a simplex in dimension $d$ into $r^d$ subsimplices, all with the same volume. We use the centroids of the subsimplices resulting from the splitting as the sampling points.

\section{Example dynamical systems}

We apply our TE estimators to time series generated by coupled instances of well known dynamical systems.
Several realizations of the time series are generated from randomly chosen initial values.  
In all cases, TE in each direction is computed using a bin size adapted to the number of points available \cite{Krakovska2018} and to the size of the reconstructed attractor (appendix \ref{app:binning}). We will generically denote the TE computed from time series $X$ and $Y$, as $TE_{X\to Y}$ and $TE_{Y\to X}$, where $X\to Y$ corresponds to the direction of the coupling (in the case of unidirectional coupling) or to the direction of the strongest coupling (in the case of bidirectional coupling). In both cases, one expects 
$TE_{X\to Y}-TE_{Y\to X} > 0$. To check the ability of our methods to detect the direction of the coupling between time series, we study the dependence of the average values of $TE_{X\to Y}$ and $TE_{Y\to X}$ across realizations, on the number of observations in the time series. We also study the response of our estimators to the strength of the coupling and to observational and dynamical noise.
We compare our results with those obtained with the kNN \cite{Kraskov2004} and the KDE \cite{steuer2002mutual} estimators.
Although these methods, strictly speaking, estimate mutual information, TE can be computed from the identity in equation (\ref{te_mi}). Because we are primarily interested in the sensitivity of our new estimators to noise and time series length, and not the absolute value of the TE, we do not apply any bias correction to the estimators \cite{Marschinski_kantz2002,Gourevitch2007,Bossomaier2016}. In appendix \ref{app:binning} we detail the embedding used for computing TE for each dynamical system example. 

\subsection{Dynamical and measurement noise}

 Measurement noise is independently added to each time series as follows: given a time series $X$, measurement noise of intensity $\epsilon$ is simulated by adding independently to each observation of $X$ a random value drawn from a Gaussian distribution with zero mean and standard deviation $\epsilon\sigma$, where $\sigma$ is the standard deviation corresponding to the values in $X$, and $0\leq \epsilon\leq 1$. For instance, a measurement noise of intensity $\epsilon = 0.1$ will be referred to as $10\%$ measurement noise. Dynamical noise is simulated for each system as explained in the following sections (equations (\ref{dyn_noise}), (\ref{Bidyn_noise}) and (\ref{RL_dyn_noise})).

\subsection{Unidirectionally coupled logistic maps (UCLM)}
      
The logistic map is one of the hallmarks of chaotic behavior in dynamical systems \cite{May1976}. It was originally proposed by R. May as a model for population growth.
 Here we consider two logistic maps unidirectionally coupled, given by 
      
     {\small \begin{align}
        x(n+1) &=  3.78\,x(n) \ls 1-x(n)\rs\,,\label{CLMx}\\
        y(n+1) &=  3.66\,f_n \ls 1-f_n\rs\,,\label{CLMy}\\
        \nonumber\\
        f_n & = \frac{y(n) + c\,x(n)}{1+c}\,. \nonumber 
      \end{align}}
     We also allow for the presence of a moderate level of dynamical noise by modifying the sequence $f_n$ as 
 {\small \begin{equation}
 \tilde f_n = \frac{y(n) + c\,(x(n) + \epsilon\,\xi)}{1+c(1+\epsilon)}\,,\label{dyn_noise}
 \end{equation}}
 where $\xi$ is a random number drawn from $[0,1]$ with a flat distribution and $0\leq \epsilon\leq 0.5$.

 \subsection{Bidirectionally coupled logistic maps (BCLM)}
 
 In this case we consider the system generated by the map
 
 {\small 
 \begin{align}
        x(n+1) &=  \frac{ 3.78\,x(n) \ls 1-x(n)\rs + 0.03\,y(n)^2}{1.03}\,,\label{BiCLMx}\\
        \nonumber\\
        y(n+1) &=  \frac{3.66\,g_n \ls 1-g_n\rs + c\,x(n)^2}{1+c}\,,\label{BiCLMy}\\
        \nonumber\\
        g_n & = \frac{y(n) + 0.06\,x(n)}{1.06}\,. \nonumber 
\end{align}}

As in the UCLM case, we also allow for the presence of dynamical noise by modifying the sequence $g_n$ as
{\small \begin{equation}
 \tilde g_n = \frac{y(n) + 0.06\,(x(n) + \epsilon\,\xi)}{1+0.06(1+\epsilon)}\,,\label{Bidyn_noise}
 \end{equation}}
 where $\xi$ is a random number drawn from $[0,1]$ with a flat distribution and $0\leq \epsilon\leq 0.5$.

\subsection{Coupled R{\"o}ssler-Lorenz system}
      
Historically relevant in the study of chaos are also the Lorenz system \cite{Lorenz1963} and the R{\"o}ssler system \cite{rossler1976equation}. The former was developed by E. Lorenz in 1963 as a simplified model of viscous fluid flow. O. R{\"o}ssler proposed his system in 1976 as a simpler version of the Lorenz attractor in order to more easily study its chaotic properties. Here we study a coupled version of both systems also studied in \cite{Krakovska2018,quiroga2000learning}, generated by the vector field 
      
      {\small \begin{align}
        \dot x_1 & = -6\,(x_2+x_3)\,,\label{RLx1}\\
        \dot x_2 & = 6\,(x_1+0.2 x_2)\,,\\
        \dot x_3 & = 6\,(0.2+x_3\,(x_1-5.7))\,,\\
        \nonumber\\
        \dot y_1 & = 10\,(y_2-y_1)\,,\\
        \dot y_2 & = y_1\,(28-y_3) - y_2 + c\,(x_2)^2\,,\label{doty2}\\
        \dot y_3 & = y_1\,y_2 - (8/3)\,y_3\,\label{RLy3}.
      \end{align}}
      
In this case, the dynamical noise is introduced by modifying the coupling term in the equation (\ref{doty2}) above as

{\small 
\begin{equation}
   c\,(x_2)^2\lb 1+\epsilon \,\xi\rb^2\,,\label{RL_dyn_noise} 
\end{equation}
} 

where $\xi$
is a random number drawn from $[-1,1]$ with a flat distribution and $0\leq \epsilon\leq 0.5$.

\subsection{Synchronization}

\begin{figure}[!ht]
	\centering

    \includegraphics[width=0.5\textwidth]{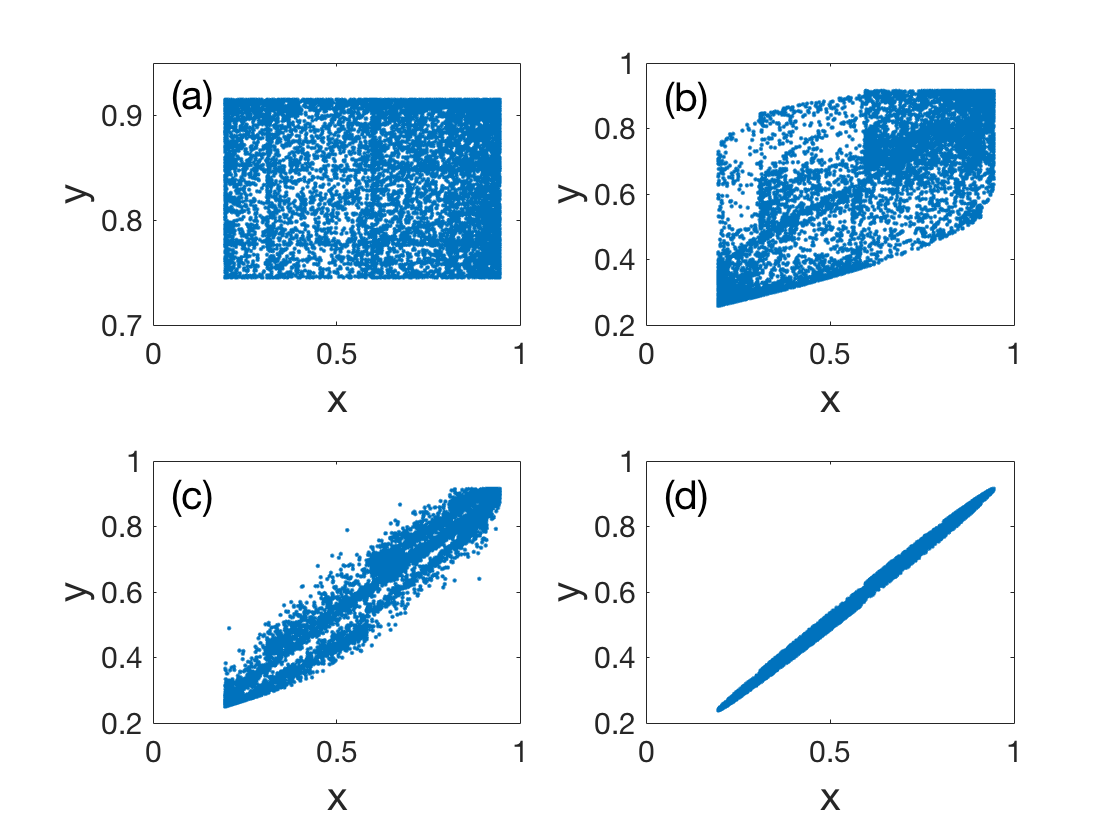}
	
	\caption{\footnotesize{$x-y$ phase space generated from $10^4$-point long orbits of the UCLM without noise and for different values of the coupling constant: $c=0$ (a), $c=0.4$ (b), $c=0.6$ (c) and $c=1$ (d).}}
    \label{fig:sync_CLM}
\end{figure}

\begin{figure}[!ht]
	\centering

    \includegraphics[width=0.5\textwidth]{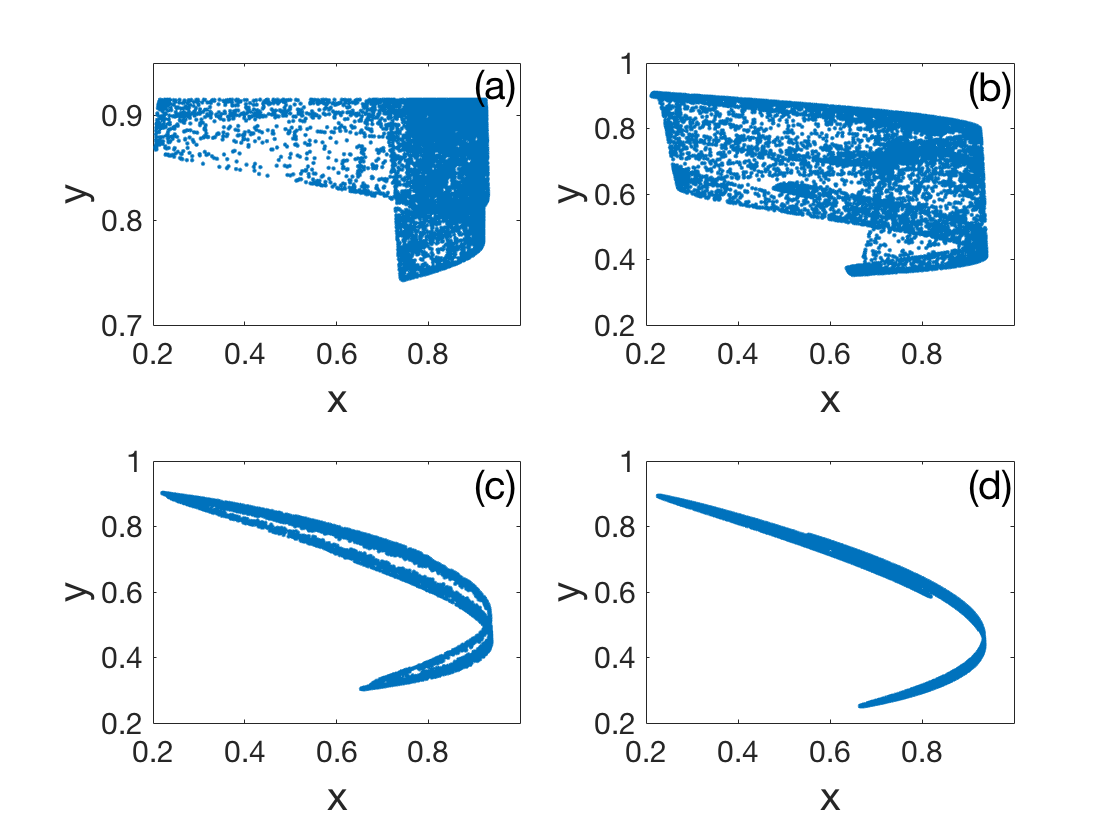}
	
	\caption{\footnotesize{$x-y$ phase space generated from $10^4$-point long orbits of the BCLM without noise and for different values of the coupling constant: $c=0$ (a), $c=0.2$ (b), $c=0.5$ (c) and $c=1$ (d).}}
    \label{fig:sync_BiCLM}
\end{figure}

The synchronization threshold for the coupled instances of the logistic maps  can be easily estimated from the shape of the attractor itself. It turns out that for both cases the synchronization seems to take place around $c\sim 1$. In the UCLM case, the attractor clearly shrinks to the diagonal for $c\sim 1$ (figure \ref{fig:sync_CLM}). For the BCLM, however, one has a generalized synchronization and hence the attractor does not collapse to the diagonal when synchronization sets in (figure \ref{fig:sync_BiCLM}). In fact, by taking the limit when $c\to\infty$, the BCLM system reduces to the new dynamical system 

{\small
\begin{align}
    x(n+1) &=  3.67\,x(n) \ls 1-x(n)\rs + 0.029\,y(n)^2\,,\label{syncBiCLMx}\\
        \nonumber\\
        y(n+1) &= x(n)^2\,.\label{syncBiCLMy}
\end{align}
}

\begin{figure}[!ht]
	\centering

    \includegraphics[width=0.5\textwidth]{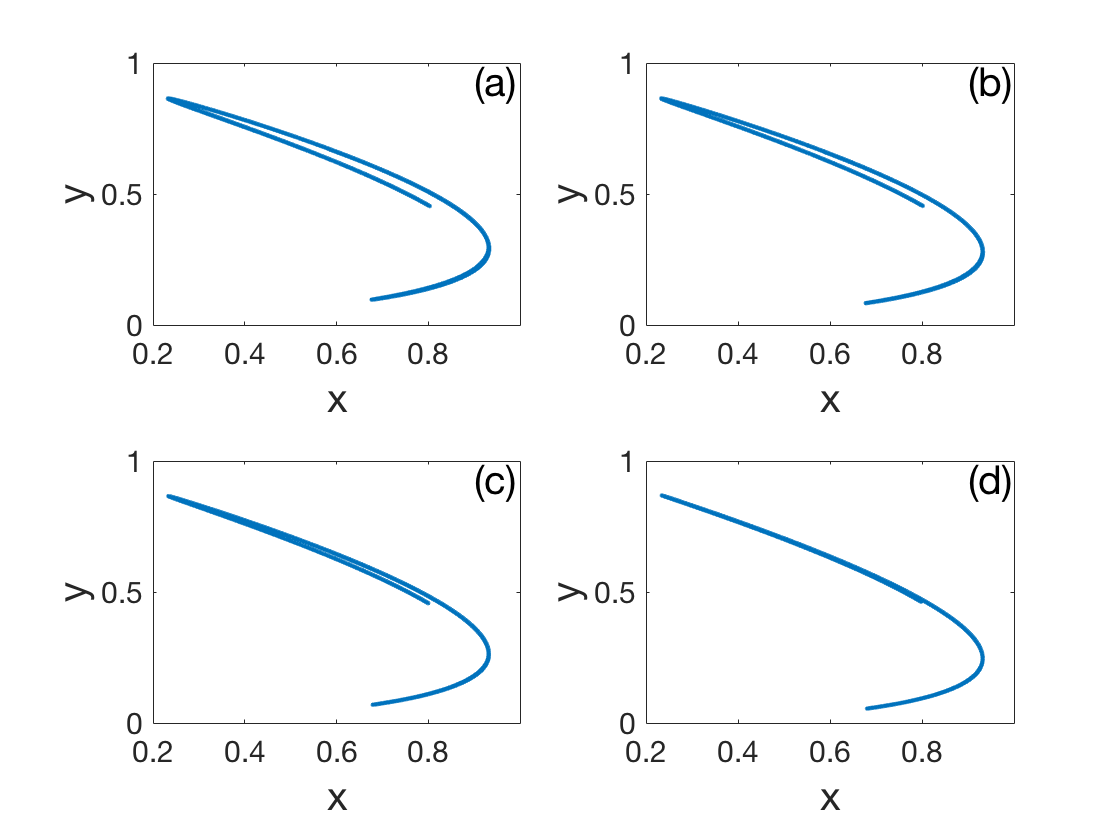}
	
	\caption{\footnotesize{
	 $x-y$ phase space generated from $10^4$-point long orbits of the BCLM without noise and for high values of the coupling constant: $c=10$ (a), $c=15$ (b), $c=30$ (c) and $c=\infty$ (the limiting system given by equations (\ref{syncBiCLMx})-(\ref{syncBiCLMy})) (d).}}
    \label{fig:BiCLM_c_infty}
\end{figure}

As $c$ increases, the attractor generated by the BCLM approaches the attractor obtained with the system (\ref{syncBiCLMx})-(\ref{syncBiCLMy}) (figure \ref{fig:BiCLM_c_infty}).

\begin{figure}[!ht]
	\centering

    \includegraphics[width=0.5\textwidth]{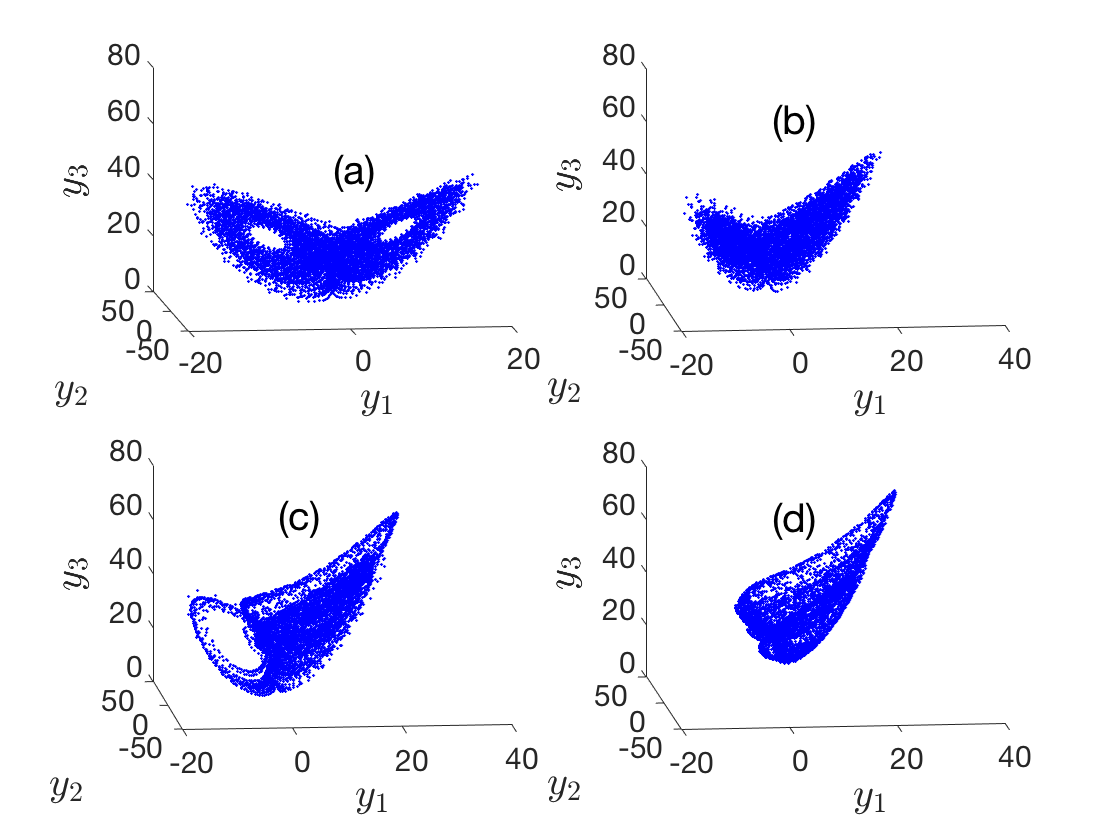}
	
	\caption{\footnotesize{3d sections of the R\"ossler-Lorenz phase space generated from $10^4$-point long orbits without noise and for different values of the coupling: $c=0$ (a), $c=1$ (b), $c=2.5$ (c) and $c=3.5$ (d).}}
    \label{fig:sync_RL}
\end{figure}

For the case of the R\"ossler-Lorenz system, the generalized synchronization seems to take place around $c\gtrsim 2$ \cite{quiroga2000learning}. Indeed, one can observe a great distortion of the usual butterfly shape Lorenz attractor for $c > 2.5$ (figure \ref{fig:sync_RL}).   

\section{Sensitivity of TE to time series length}

In this section we study the dependence of the TE computed using our estimators on the number of observations in the time series. 

\subsection{Coupled logistic maps}

\subsubsection{Data-rich time series}
      
In both UCLM and BCLM systems, we compute the TE as a function of the time series length in the range $1000$ to $5000$ observations and with a low level ($10\%$) of measurement noise. For the UCLM case the coupling constant is set to $c = 0.4$, while for the BCLM instance we set $c = 0.2$.  

\begin{figure}[!ht]
     %\vspace{-3.5cm}
  \includegraphics[width=0.5\textwidth]{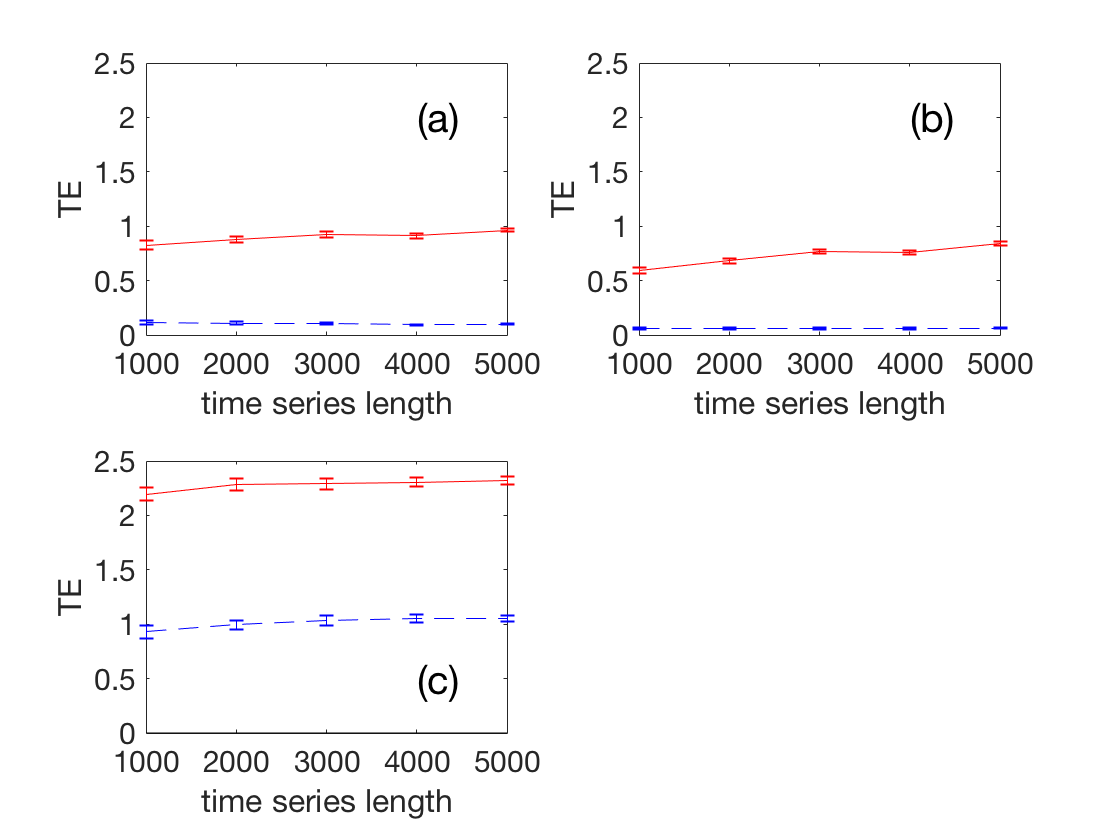} 
    %\vspace{-3.5cm}
	\caption{\footnotesize{
	Sensitivity of $TE$ to time series length for the UCLM with coupling constant $c = 0.4$ and 10\% measurement noise, using data rich time series. Values are the mean and standard deviation of $TE_{x\to y}$ (red line) and $TE_{y\to x}$ (dashed blue line) over 50 realizations, computed with the grid estimator (a), the KDE estimator (b) and the kNN estimator (c). }}
    \label{fig:UCLM_thousands_tsl}
    
\end{figure}

\begin{figure}[!ht]
     %\vspace{-3.5cm}
  \includegraphics[width=0.5\textwidth]{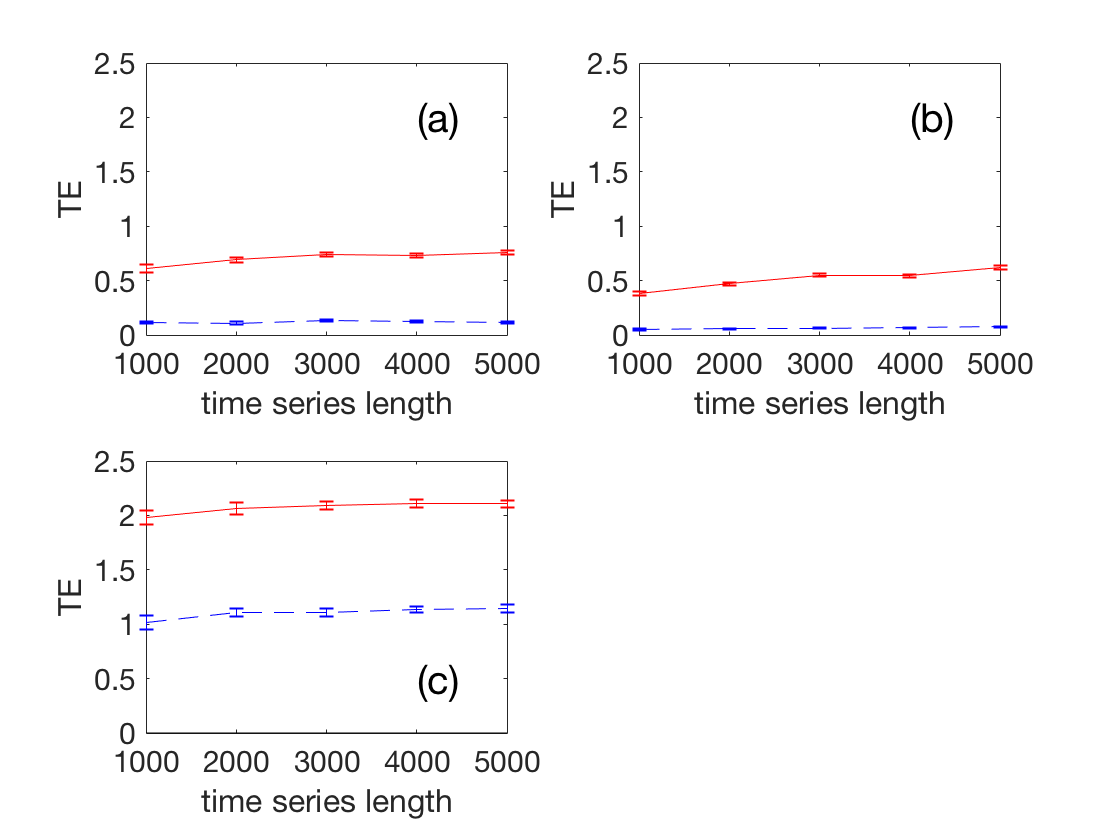} 
    %\vspace{-3.5cm}
	\caption{\footnotesize{
	Sensitivity of $TE$ to time series length for the BCLM with coupling constant $c = 0.2$ and 10\% measurement noise, using data rich time series. Values are the mean and standard deviation of of $TE_{x\to y}$ (red line) and $TE_{y\to x}$ (dashed blue line) over 50 realizations, computed with the grid estimator (a), the KDE estimator (b) and the kNN estimator (c).  
	}}
    \label{fig:BCLM_thousands_tsl}
    
\end{figure}

In both cases of coupled logistic maps, the grid and the kNN estimators seem to be the least sensitive to the number of observations in the time series (figures \ref{fig:UCLM_thousands_tsl}a,c and \ref{fig:BCLM_thousands_tsl}a,c), while the KDE estimator shows a mildly stronger dependence on the time series length. For the grid estimator, and using the adapted bin size described in appendix \ref{app:binning}, TE for both UCLM and BCLM saturates to a fixed value for time series with more than $\sim 5000$ observations.

 \subsubsection{Sparse time series}
 
\begin{figure}[!ht]
      \centering
      
      \includegraphics[width=0.5\textwidth]{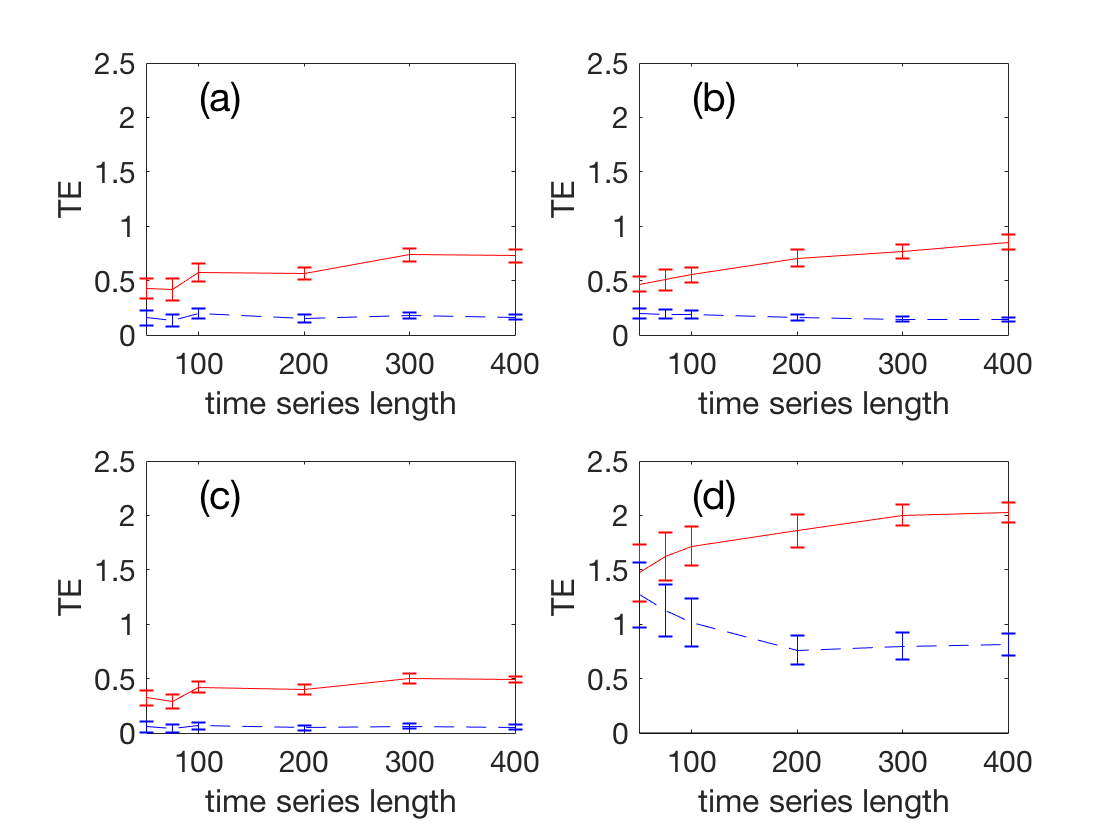}
     
	\caption{\footnotesize{
	Sensitivity of $TE$ to time series length for the UCLM with coupling constant $c = 0.4$ and 10\% measurement noise, using sparse time series.
	Values are the mean and standard deviation of of $TE_{x\to y}$ (red line) and $TE_{y\to x}$ (dashed blue line) over 50 realizations, computed with the grid estimator (a), the triangulation estimator (b), the KDE estimator (c) and the kNN estimator (d).}}
    \label{fig:UCLM_hundreds_tsl}
    
\end{figure}

\begin{figure}[!ht]
      \centering
      
      \includegraphics[width=0.5\textwidth]{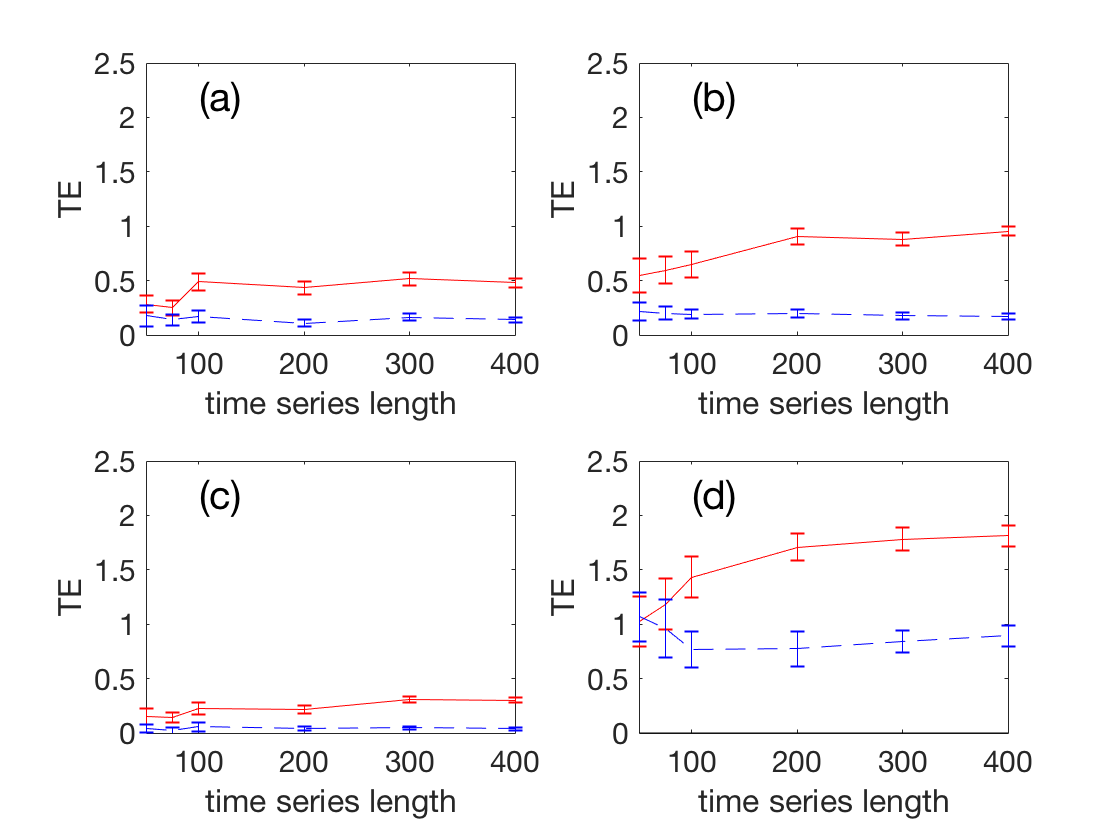}
     
	\caption{\footnotesize{
	Sensitivity of $TE$ to time series length for the BCLM with coupling constant $c = 0.2$ and 10\% measurement noise, using sparse time series. Values are the mean and standard deviation of of $TE_{x\to y}$ (red line) and $TE_{y\to x}$ (dashed blue line) over 50 realizations, computed with the grid estimator (a), the triangulation estimator (b), the KDE estimator (c) and the kNN estimator (d).}}
    \label{fig:BCLM_hundreds_tsl}
    
\end{figure}

  In this section we check the ability of our estimators to yield directional asymmetry in the TE for sparse data in the range 50 to 400 observations and adding 10\% measurement noise. We also test the stability of the TE against the time series length. For both UCLM and BCLM systems, our estimators yield the correct TE asymmetry ($TE_{x\to y} > TE_{y\to x}$) even for time series sparsely sampled with 50 values (figures \ref{fig:UCLM_hundreds_tsl}a,b  and \ref{fig:BCLM_hundreds_tsl}a,b). In the case of the BCLM system, the triangulation estimator outcompetes the rest of the estimators at detecting asymmetry in the TE (in the expected direction) for very sparse time series, with less than 100 observations (figure \ref{fig:BCLM_hundreds_tsl}).
 As for the sensitivity of the TE on the number of observations, our estimators yield relatively stable TE in the range $100-300$ observations (figures \ref{fig:UCLM_hundreds_tsl}a,b and \ref{fig:BCLM_hundreds_tsl}a,b). The KDE estimator seems to be the least sensitive to the number of observations while the kNN estimator shows the highest sensitivity (figures \ref{fig:UCLM_hundreds_tsl}c,d and \ref{fig:BCLM_hundreds_tsl}c,d).

%\newpage
\subsection{Coupled R{\"o}ssler-Lorenz system}

\begin{figure}[!ht]
	\centering
    
    %\vspace{0.1cm}
    \includegraphics[width=0.5\textwidth]{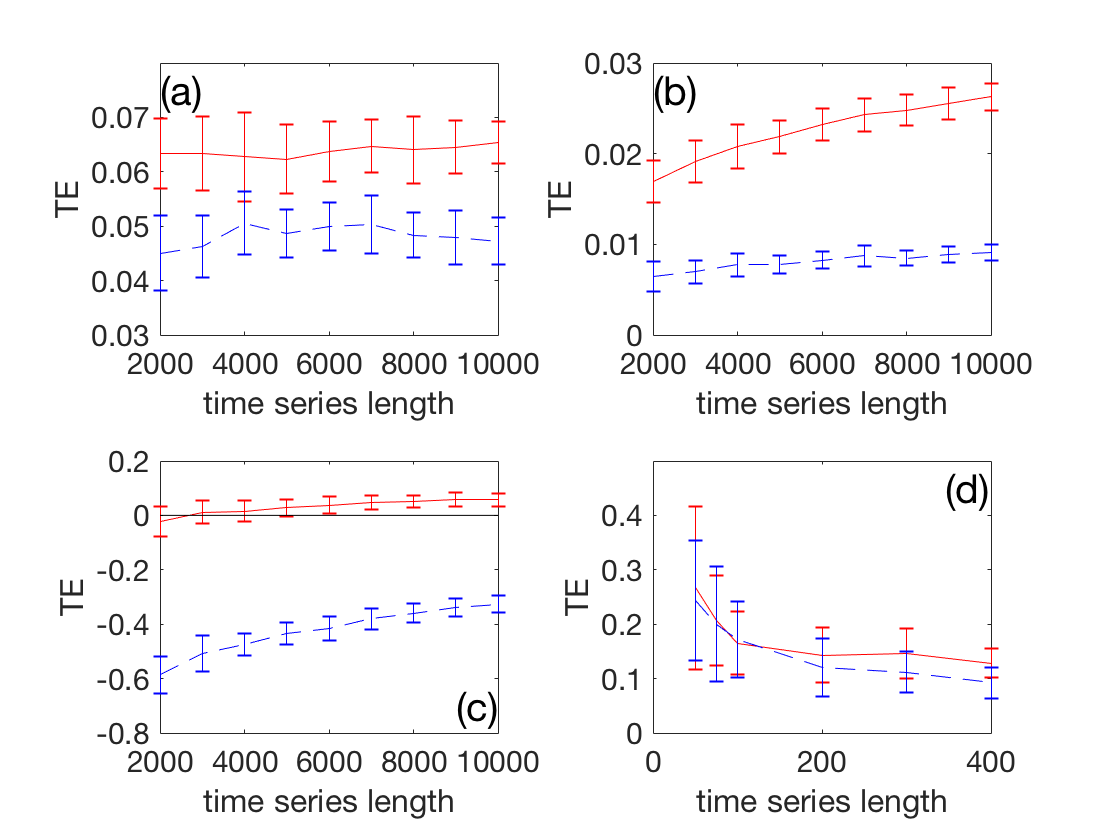}
    
    %\vspace{-1.5cm}
    
	\caption{\footnotesize{Sensitivity of $TE$ to time series length for the R\"ossler-Lorenz system with coupling constant $c=1.5$ and 10\% measurement noise. 
	Values are the mean and standard deviation of $TE_{x_2\to y_2}$ (red line) and $TE_{y_2\to x_2}$ (dashed blue line) over 50 realizations, computed with the grid estimator (a), the KDE estimator (b) and the kNN estimator (c). We also include the TE computed using the triangulation estimator with 3d embeddings and using sparse time series (d).}}
    \label{fig:RL_tsl}
\end{figure}

\noindent  Due to the dimensionality of this system, relatively data-rich time series are required to obtain reliable TE estimates, hence we use time series with 2000 to 10000 observations. Again we compare the TE computed using our grid transfer operator estimator with that of the kNN and the KDE estimators. The triangulation approach becomes prohibitively time-demanding for high embedding dimension, $d\geq 5$ (appendix \ref{app:timing}). 
The grid estimator is the least sensitive to the number of observations in the time series (figure \ref{fig:RL_tsl}a) while the kNN estimator arguably is the most sensitive (figure \ref{fig:RL_tsl}c). 
For the grid estimator, and using the adapted bin size described in appendix \ref{app:binning}, TE for the R\"ossler-Lorenz system saturates to a fixed value for time series with more than $\sim 12000$ observations.
For completeness, we also applied the triangulation estimator to the R\"ossler-Lorenz system using $3$d embeddings ($(x_2(i+1),x_2(i),y_2(i))$, for computing $TE_{y_2\to x_2}$ and $(y_2(i+1),y_2(i),x_2(i))$, for computing $TE_{x_2\to y_2}$) and time series with 50-400 observations. The estimator detects a marginal causal signal for time series with more than $\sim 300$ observations (figure \ref{fig:RL_tsl}d).

We note that the computational requirements of the triangulation estimator become prohibitive for embedding dimensions beyond $\sim 5$.  The current implementation of this estimator is based on computing exact simplex volume intersections (appendix \ref{app:PolytopeVol}) which suffers from the curse of dimensionality. There is room for optimization, however, through more efficient (approximate) polytope intersection algorithms.
Alternatively, a coarse grained sampling of the simplices might also be used to estimate the volume intersection. We have implemented both exact and approximate volume intersection routines in our CausalityTools.jl \footnote{CausalityTools.jl is a registered Julia package and the source code is found at \href{https://github.com/kahaaga/CausalityTools.jl}{https://github.com/kahaaga/CausalityTools.jl}} Julia \cite{Bezanson2017} package, which also provides an implementation of the grid estimator. We also suggest a more efficient alternative in appendix \ref{app:mixed}. We leave a comprehensive investigation of these optimizations for future work.

\section{Dependence of TE on the coupling constant and its response to noise}

In this section we study the dependence of the TE computed using our estimators on the coupling constant and its response to observational and dynamical noise. We compare the results with the above standard estimators. Because our main interest in this study is the estimation of TE from sparse time series, we fix the time series length to 100 observations for the coupled logistic maps (both UCLM and BCLM) and to 1000 observations for the the R\"ossler-Lorenz system.

\subsection{Coupled logistic maps}

\begin{figure}[!ht]
	\centering
    
    %\vspace{0.1cm}
    \includegraphics[width=0.5\textwidth]{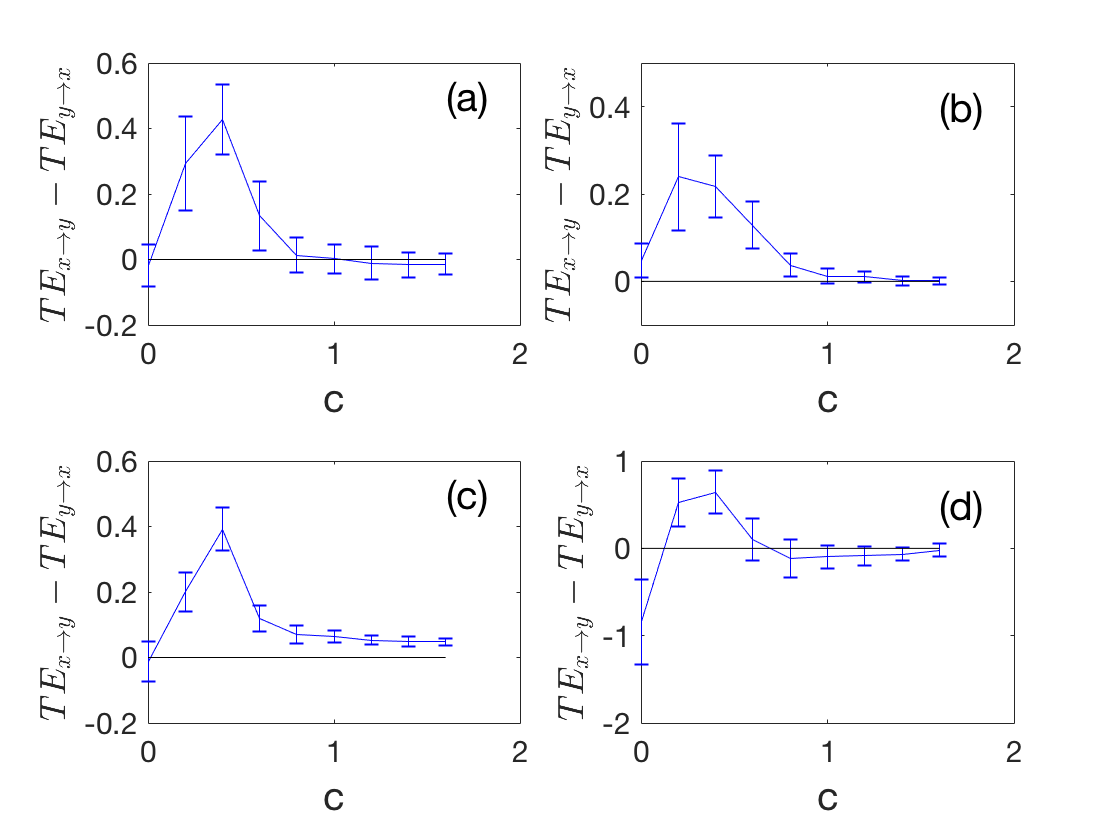}
    
    %\vspace{-1.5cm}
    
	\caption{\footnotesize{Dependence of $TE_{x\to y}-TE_{y\to x}$ on the coupling constant for the noise-free UCLM. Values are the mean and standard deviation of $TE_{x\to y}-TE_{y\to x}$ over 50 realizations, computed with the grid estimator (a), the triangulation estimator (b), the KDE estimator (c) and the kNN estimator (d).}}
    \label{fig:UCLM_c_noisless}
\end{figure}

\begin{figure}[!ht]
	\centering
    
    %\vspace{0.1cm}
    \includegraphics[width=0.5\textwidth]{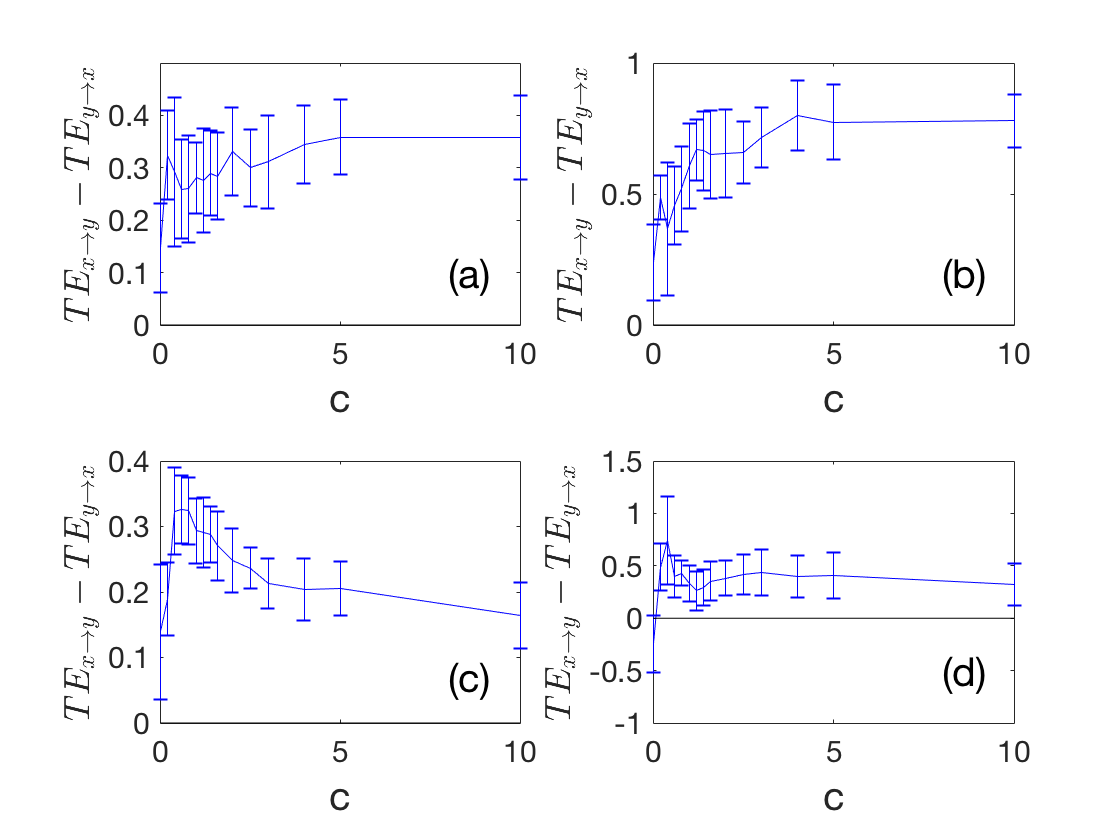}
    
    %\vspace{-1.5cm}
    
	\caption{\footnotesize{Dependence of $TE_{x\to y}-TE_{y\to x}$ on the coupling constant for the noise-free BCLM.  Values are the mean and standard deviation of $TE_{x\to y}-TE_{y\to x}$ over 50 realizations, computed with the grid estimator (a), the triangulation estimator (b), the KDE estimator (c) and the kNN estimator (d).}}
    \label{fig:BCLM_c_noisless}
\end{figure}

\begin{figure}[!ht]
	\centering
    
    %\vspace{0.1cm}
    \includegraphics[width=0.5\textwidth]{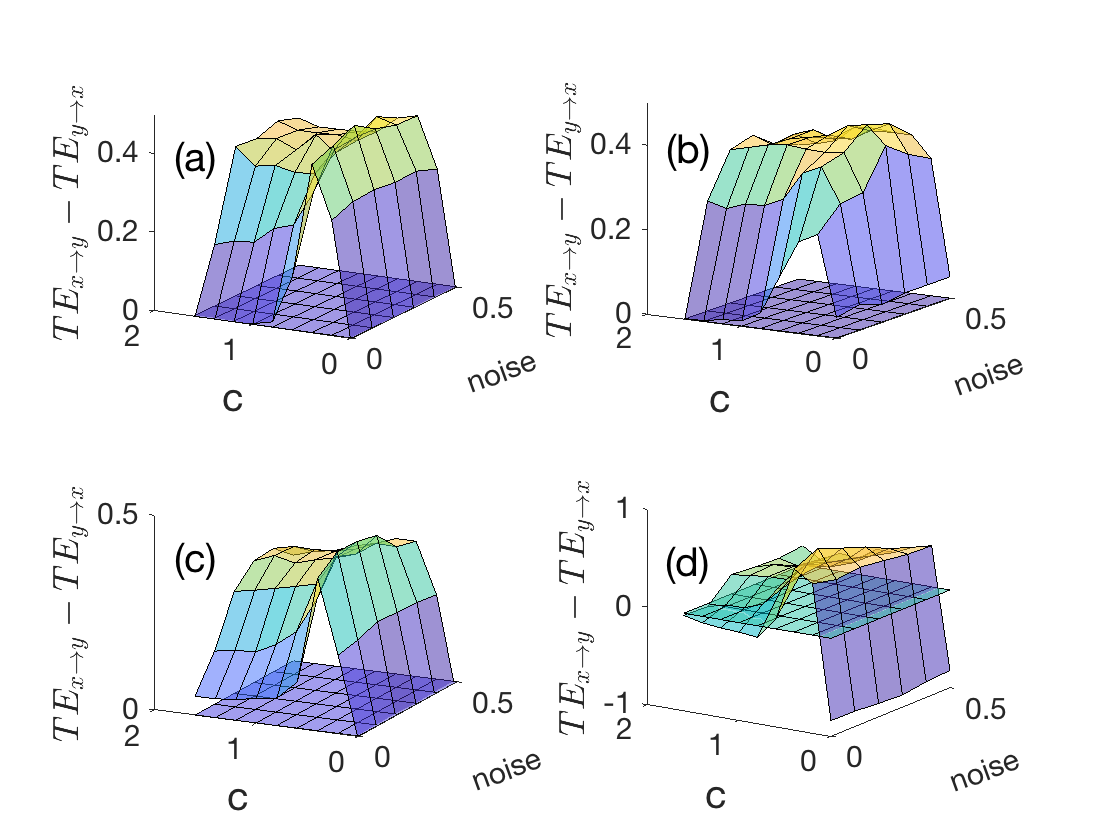}
    
    %\vspace{-1.5cm}
    
	\caption{\footnotesize{Dependence of $TE_{x\to y}-TE_{y\to x}$ on the coupling constant and dynamical noise level for the UCLM. Values are the mean and standard deviation of $TE_{x\to y}-TE_{y\to x}$ over 50 realizations, computed with the grid estimator (a), the triangulation estimator (b), the KDE estimator (c) and the kNN estimator (d).}}
    \label{fig:UCLM_c_dyn_noise}
\end{figure}

\begin{figure}[!ht]
	\centering
    
    %\vspace{0.1cm}
    \includegraphics[width=0.5\textwidth]{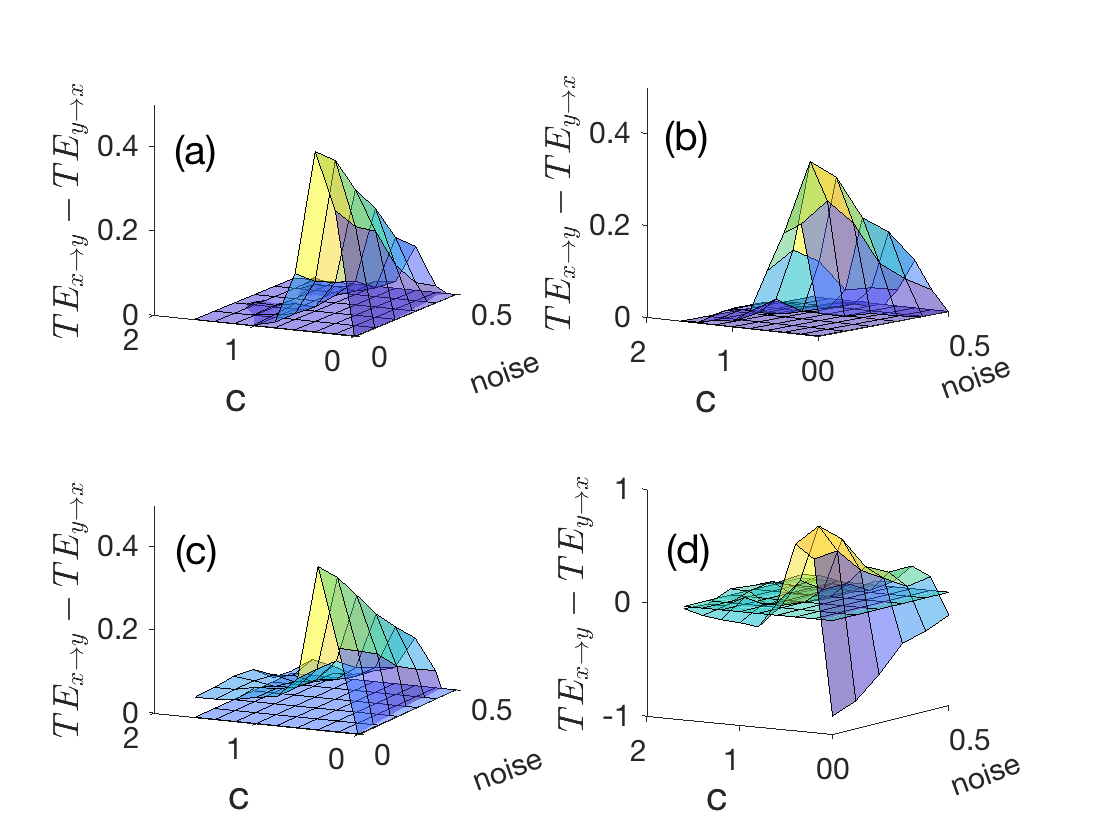}
    
    %\vspace{-1.5cm}
    
	\caption{\footnotesize{Dependence of $TE_{x\to y}-TE_{y\to x}$ on the coupling constant and measurement noise level for the UCLM. Values are the mean and standard deviation of $TE_{x\to y}-TE_{y\to x}$ over 50 realizations, computed with the grid estimator (a), the triangulation estimator (b), the KDE estimator (c) and the kNN estimator (d).}}
    \label{fig:UCLM_c_meas_noise}
\end{figure}

\begin{figure}[!ht]
	\centering
    
    %\vspace{0.1cm}
    \includegraphics[width=0.5\textwidth]{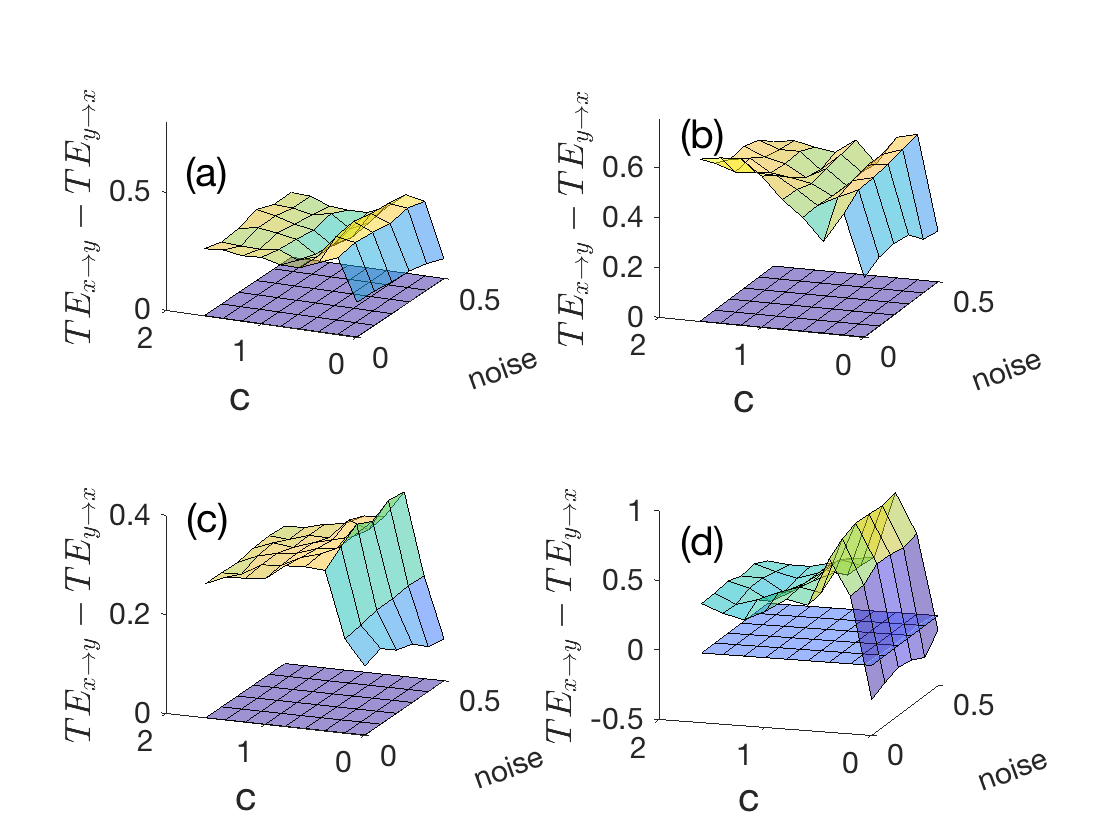}
    
    %\vspace{-1.5cm}
    
	\caption{\footnotesize{Dependence of $TE_{x\to y}-TE_{y\to x}$ on the coupling constant and dynamical noise level for the BCLM. Values are the mean and standard deviation of $TE_{x\to y}-TE_{y\to x}$ over 50 realizations, computed with the grid estimator (a), the triangulation estimator (b), the KDE estimator (c) and the kNN estimator (d).}}
    \label{fig:BCLM_c_dyn_noise}
\end{figure}

\begin{figure}[!ht]
	\centering
    
    %\vspace{0.1cm}
    \includegraphics[width=0.5\textwidth]{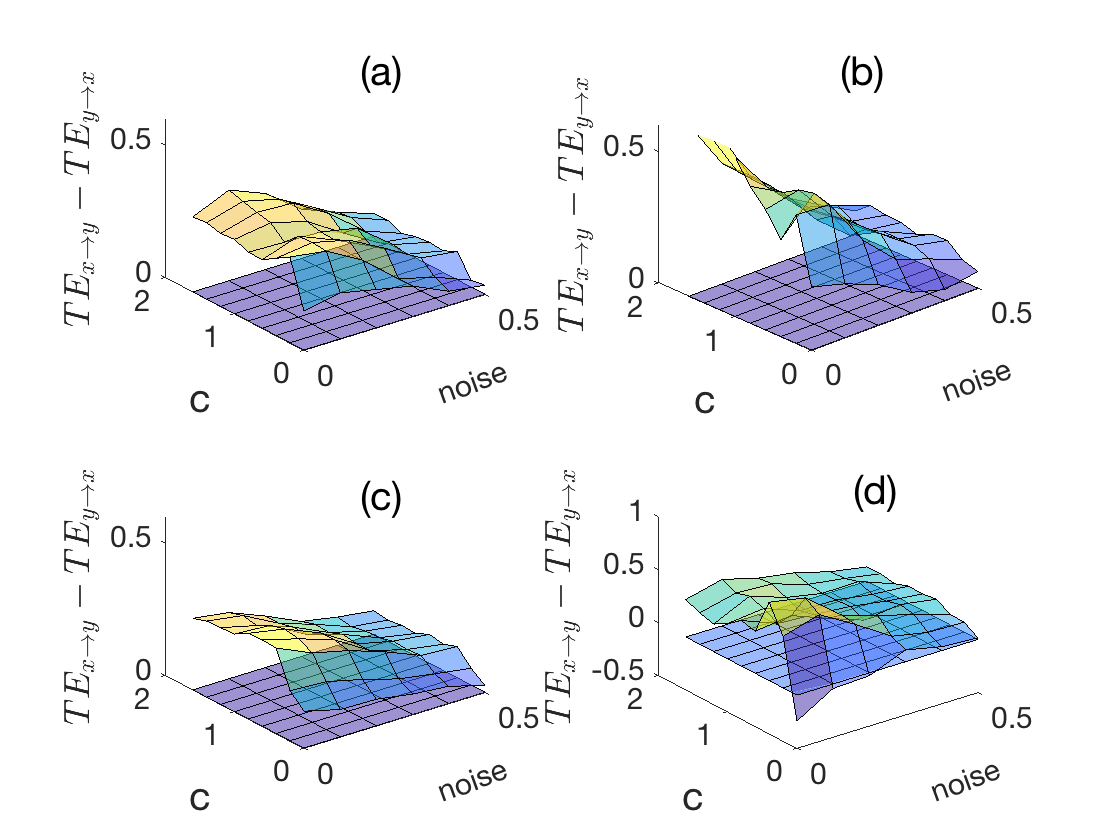}
    
    %\vspace{-1.5cm}
    
	\caption{\footnotesize{Dependence of $TE_{x\to y}-TE_{y\to x}$ on the coupling constant and measurement noise level for the BCLM. Values are the mean and standard deviation of $TE_{x\to y}-TE_{y\to x}$ over 50 realizations, computed with the grid estimator (a), the triangulation estimator (b), the KDE estimator (c) and the kNN estimator (d).}}
    \label{fig:BCLM_c_meas_noise}
\end{figure}

For the coupled logistic maps we compute the TE for values of the coupling constant in the range $0$ to $1.6$ in steps of $0.2$. We also include dynamical noise as well as measurement noise with intensities ranging from $0$ to $0.5$ in steps of $0.1$. For the case of UCLM without noise, the asymmetry $TE_{x\to y} - TE_{y\to x}$, computed with our estimators, starts at zero (or a very small value) for $c=0$, it then increases up to a maximum value around $c \sim 0.2-0.4$ and decreases back to zero as the synchronization triggers for $c\gtrsim 1$ (figures \ref{fig:UCLM_c_noisless}a,b). In the UCLM system, synchronization causes the evolution of $y$ to closely follow the evolution of $x$ (figure \ref{fig:sync_CLM}). Accordingly, one expects $TE_{x\to y} \to TE_{y\to x}$ as $c$ increases beyond $1$. 
%It also seems that the triangulation estimator reacts more sensitively to the synchronization compared with the grid estimator and, especially, with the KDE estimator (figures \ref{fig:UCLM_c_noisless}a, \ref{fig:UCLM_c_noisless}b and \ref{fig:UCLM_c_noisless}c).
However, when dynamical noise is added, the effect of the synchronization is attenuated (figures \ref{fig:UCLM_c_dyn_noise}a,b). Dynamical noise may be interpreted as a hidden process affecting the system. It is then expected that the synchronization effect breaks down for sufficiently intense dynamical noise. On the other hand, when measurement noise is added, the asymmetry $TE_{x\to y} - TE_{y\to x}$ decreases with increasing noise intensity (figure \ref{fig:UCLM_c_meas_noise}), as the effect of the coupling is masked by the noise. Remarkably, the ability of the triangulation estimator to detect the correct directionality of the coupling enhances for low to moderate levels of measurement noise and weak coupling ($c\sim 0.2-0.4$) (figure \ref{fig:UCLM_c_meas_noise}b). This finding is congruent with the concept of random perturbation approximation to the map, on which the triangulation estimator is based \cite{Froyland1997}.
In the case of BCLM, synchronization does not decrease the asymmetry between $TE_{x\to y}$ and $TE_{y\to x}$. Synchronization reduces the BCLM system to the map in equations (\ref{syncBiCLMx})-(\ref{syncBiCLMy}). 
In that limiting case, the coupling in the $x\to y$ direction is much stronger than in the opposite direction. Both the grid and the triangulation estimators do capture this saturation of the asymmetry in the TE for high values of the coupling constant (figures \ref{fig:BCLM_c_noisless}a,b). In contrast, the KDE estimator yields a monotonically decreasing asymmetry $c\gtrsim 1$ (figure \ref{fig:BCLM_c_noisless}c) whereas the kNN estimator shows a less obvious decrease (figure \ref{fig:BCLM_c_noisless}d). When dynamical noise is included into the BCLM system, the asymmetry $TE_{x\to y}-TE_{y\to x}$ responds similarly to that of the UCLM system (figure \ref{fig:BCLM_c_dyn_noise}). Observational noise, however, causes the TE asymmetry to decreases with increasing levels of noise, as expected. (figure \ref{fig:BCLM_c_meas_noise}). 
Although less obvious than for the UCLM system, the triangulation estimator also shows a local maximum in TE asymmetry for weak coupling and moderate observational noise (figure \ref{fig:BCLM_c_meas_noise}b).

\subsection{R\" ossler-Lorenz system}

\begin{figure}[!ht]
	\centering
    
    %\vspace{0.1cm}
    \includegraphics[width=0.5\textwidth]{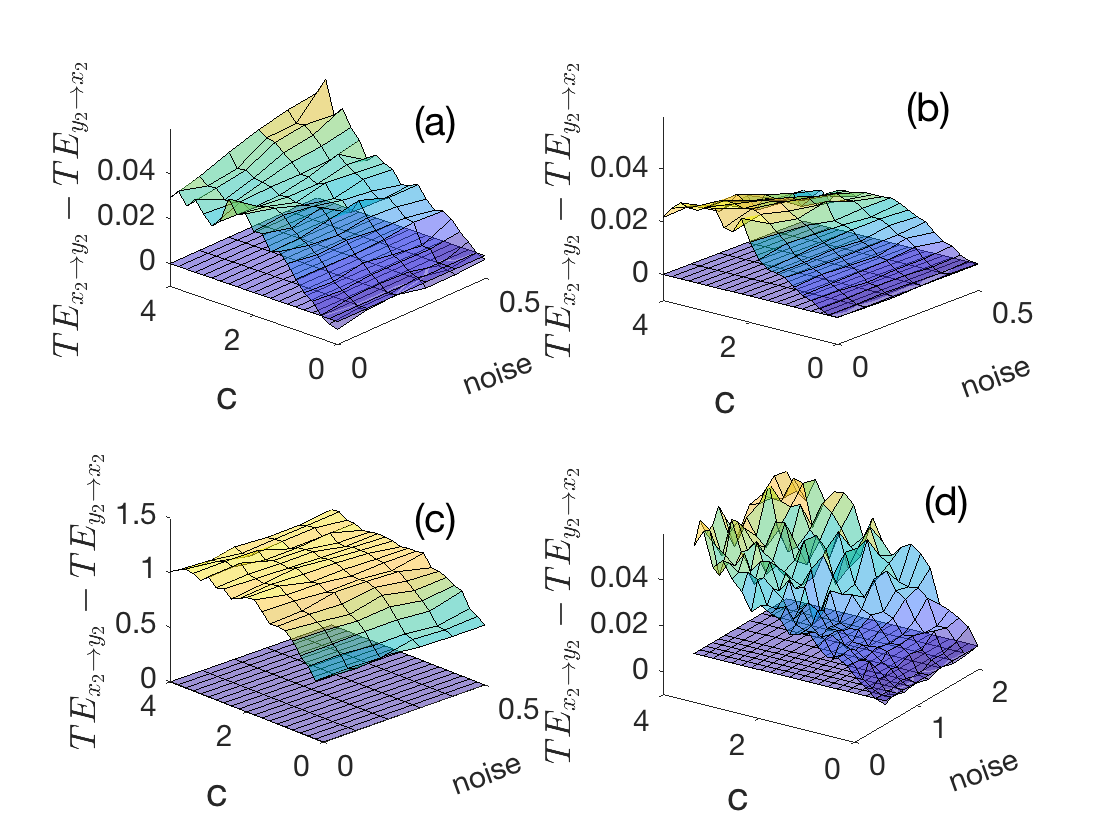}
    
    %\vspace{-1.5cm}
    
	\caption{\footnotesize{Dependence of $TE_{x_2\to y_2}-TE_{y_2\to x_2}$ on the coupling constant and dynamical noise level for the R\"ossler-Lorenz system. (a) Mean value of $TE_{x_2\to y_2}-TE_{y_2\to x_2}$ over 50 realizations computed with the grid estimator; (b) the same for the KDE estimator; (c) using the kNN estimator; (d) using the grid estimator and extending level of dynamical noise up to 2.}}
    \label{fig:RL_c_dyn_noise}
\end{figure}

\begin{figure}[!ht]
	\centering
    
    %\vspace{0.1cm}
    \includegraphics[width=0.5\textwidth]{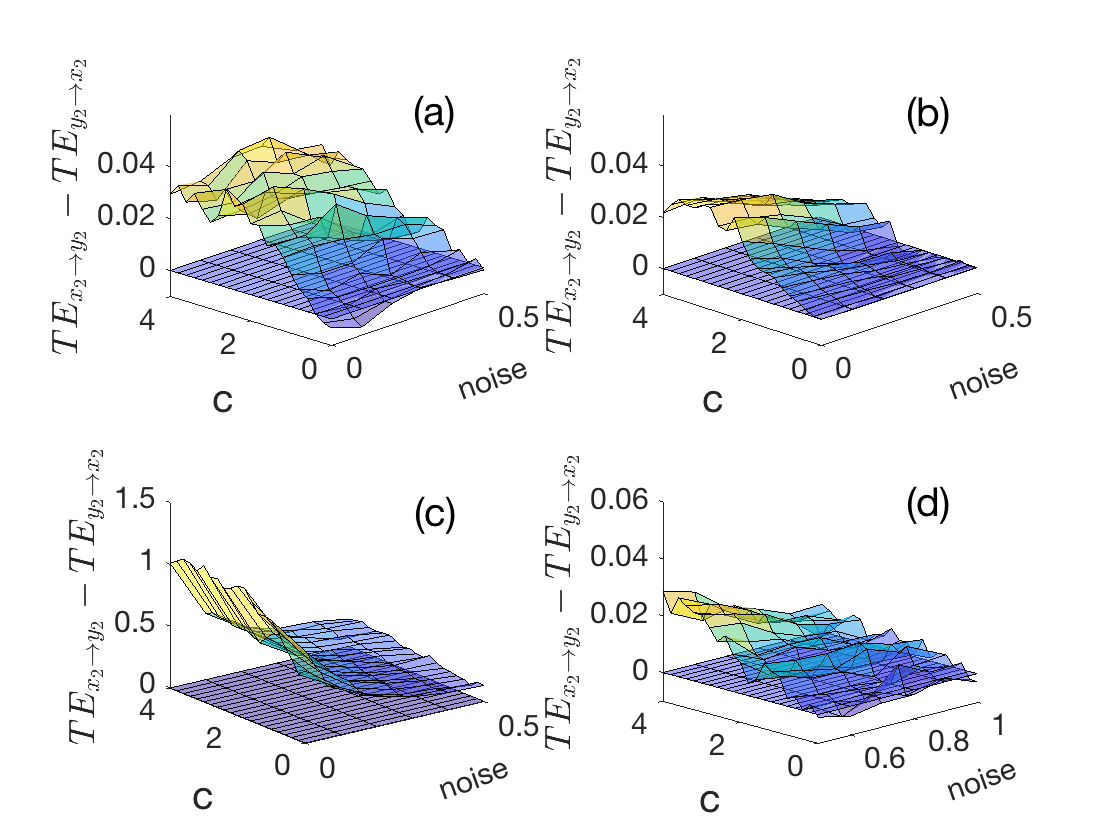}
    
    %\vspace{-1.5cm}
    
	\caption{\footnotesize{Dependence of $TE_{x_2\to y_2}-TE_{y_2\to x_2}$ on the coupling constant and measurement noise level for the R\"ossler-Lorenz system. (a) Mean value of $TE_{x_2\to y_2}-TE_{y_2\to x_2}$ over 50 realizations computed with the grid estimator; (b) the same for the KDE estimator; (c) using the kNN estimator; (d) using the grid estimator and extending the level of measurement noise up to 100\%.}}
    \label{fig:RL_c_meas_noise}
\end{figure}

For the R\" ossler-Lorenz system, we compute the TE for values of the coupling constant in the range $0$ to $4$ in steps of $0.2$ also including observational and dynamical noise with intensities ranging from $0$ to $0.5$ in steps of $0.1$. The asymmetry $TE_{x_2\to y_2}-TE_{y_2\to x_2}$ computed with the grid estimator saturates for high values of dynamical noise and coupling constant (figure \ref{fig:RL_c_dyn_noise}d), a trait also seen for the kNN estimator (figure \ref{fig:RL_c_dyn_noise}c). The TE asymmetry computed using the grid estimator increases for strong coupling and high levels of dynamical noise (figure \ref{fig:RL_c_dyn_noise}a). However, the asymmetry saturates for higher levels of dynamical noise (figure \ref{fig:RL_c_dyn_noise}d).
With increasing levels of measurement noise, all estimators yield a decreasing value for the TE asymmetry (figure \ref{fig:RL_c_meas_noise}). Remarkably, nonetheless, the TE asymmetry computed with the grid estimator only starts to decrease substantially once the level of measurement noise goes beyond $50\%$ (figure \ref{fig:RL_c_meas_noise}d). These findings suggest that our grid estimator is robust to both observational and dynamical noise.

%\newpage

\section{Direct vs indirect coupling}\label{condTE}

At the end of section \ref{sec:te_comp}, we mentioned that our approach to estimate TE can be easily extended to compute conditional transfer entropy between 3 time series. As an example of this, we apply the grid estimator to a chain of three coupled Lorenz systems, studied in \cite{kugiumtzis2013direct}. The flow is generated by the vector field

\begin{figure}[!ht]
	\centering
    
    %\vspace{0.1cm}
    \includegraphics[width=0.5\textwidth]{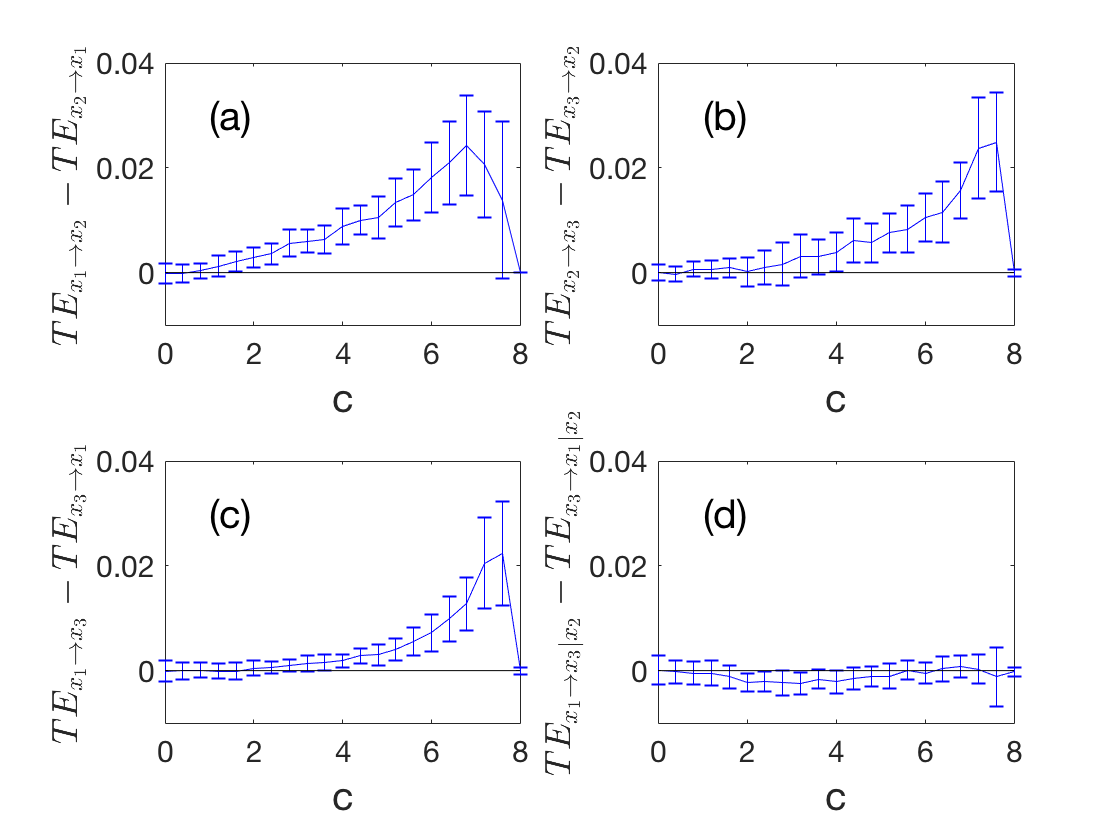}
    
    %\vspace{-1.5cm}
    
	\caption{\footnotesize{Distinguishing direct from indirect coupling for the chain of coupled Lorenz systems. Values are the mean and standard deviation of TE asymmetry over 50 realizations of the chain of coupled Lorenz systems (equations (\ref{eq1})-(\ref{eq2})). TE asymmetry for the direct coupling $x_1\to x_2$ (a), direct coupling $x_2\to x_3$ (b), indirect coupling $x_1\to x_3$ (c) and conditional TE for the indirect coupling given the mediating variable (d).}}
    \label{fig:LorenzChain}
\end{figure}

{\small 
\begin{align}
    \dot x_1 &= 10 (y_1 - x_1)\,,\label{eq1}\\
    \dot y_1 &= x_1(28-z_1) - y_1\,,\\
    \dot z_1 & = x_1 y_1 - 8/3\,z_1\,,\\
    \nonumber\\
    \dot x_i &= 10 (y_i - x_i) + c \,(x_{i-1}-x_i)\,,\\
    \dot y_i &= x_i(28-z_i) - y_i\,,\\
    \dot z_i & = x_i y_i - 8/3\,z_i\,,\label{eq2}
\end{align}
}

with $i=2,3$. The direct coupling chain is $x_1\to x_2\to x_3$. We use coupling constant values in the range $0$ to $8$ in steps of $0.4$ (according to \cite{kugiumtzis2013direct}, the full synchronization takes place for $c > 8$). For each instance of the coupling constant, we generate 50 orbits starting at randomly chosen initial conditions and consisting of $10^4$ observations. The data generation for this system and the delay embeddings used to compute TE are specified in Appendix \ref{app:binning}.
%Each orbit is obtained by solving the system of ODEs using an order 4 Runge-Kutta method with time step $dt = 0.005$. The time series are generated by sampling the variables $x_1,x_2$ and $x_3$ every 6 integration steps and after a lapse of 500 integration steps.
%To compute $TE_{x_i\to x_j}$, with $i\neq j\in \lc 1,2,3\rc$, we use the embedding $(x_j(t+3),x_j(t+2),x_j(t+1),x_i(t+2),x_i(t+1),x_i(t))$ while for the estimation of the conditional transfer entropy $TE_{x_1\to x_3|x_2}$ we used the embedding $(x_3(t+3),x_3(t+2),x_1(t+2),x_2(t+2),x_2(t+1),x_2(t))$. For the case 
%$TE_{x_3\to x_1|x_2}$, the embedding is obtained as in the case of $TE_{x_1\to x_3|x_2}$ by simply interchanging the roles of $x_3$ and $x_1$.
Our grid estimator detects the direct coupling $x_1\to x_2$ for $c \gtrsim 2$ (figure \ref{fig:LorenzChain}a) and the coupling $x_2\to x_3$ is detected for $c\gtrsim 4$ 
(figure \ref{fig:LorenzChain}b). In addition, the grid estimator detects the indirect coupling $x_1\to x_3$ for $c\gtrsim 4$ (figure \ref{fig:LorenzChain}c). On the other hand, when the transfer entropy is conditioned on the mediating variable $x_2$, the TE for the indirect coupling $x_1\to x_3$ vanishes (figure \ref{fig:LorenzChain}d),
which indicates that our method holds some promise for detecting indirect coupling.

%\newpage

\section{Conclusions}

In this work we propose the computation of transfer entropy (TE) between time series corresponding to variables of some dynamical system, based on a numerical approximation of the Perron-Frobenius operator (transfer operator) associated to the map (or vector field) giving rise to the dynamics. More specifically, the TE is computed using the invariant distribution of the transfer operator. Depending on the number of observations in the time series and the embedding dimension, we propose two methods to estimate the transfer operator. For sparse time series (less than a few hundred points) and low embedding dimension, we use a triangulation of the delay reconstructed attractor to estimate the transfer operator, whereas for data-rich time series (thousands of points) or high embedding dimension we estimate TE using a faster rectangular grid approach.
The TE computed using our estimators shows robustness to both observational and dynamical noise, even for a high dimensional system such as the R\"ossler-Lorenz system.
%We yet propose an alternative approach that lies within these two, described as a mixed approach, where from an initial triangulation generated out of few hundred points, one produces thousands of points (as a result of the sampling of each simplex) for which their images are computed by assuming a piecewise linear approximation of the underlying map.  
%This intermediate approach enables reliable estimates of the TE from sparse time series.
Our results suggest that our estimators of TE are relevant for the detection of causal directionality between sparse and noisy time series, which are commonly encountered in many disciplines. \\

\vspace{1cm}

Computer code required for reproducing the numerical results presented in this work is available in our CausalityTools.jl Julia package, for which the source code is found at \href{https://github.com/kahaaga/CausalityTools.jl}{https://github.com/kahaaga/CausalityTools.jl}.

\section*{Acknowledgements} % not compulsory

%Acknowledgements should be brief, and should not include thanks to anonymous referees and editors, or effusive comments. Grant or contribution numbers may be acknowledged.

This work has been funded by the Bergen Research Foundation  and by the Norwegian Research Council grant no. 231259.

\appendix

\section{Numerical implementation details.}
\label{app:binning}

\subsection{Generating time series and embedding}

%\begin{enumerate}
\subsubsection{Logistic maps}

To generate the time series for the UCLM and BCLM systems,
the variables $x$ and $y$ are sampled every second iterate, after a lapse of $10^3$ iterations. For the computation of $TE_{x\to y}$, we used the embedding $(y(i +1) , y(i) , x(i))$ while the computation of $TE_{y\to x}$ was done with the embedding $(x(i+1) , x(i) , y(i))$.

\subsubsection{R{\"o}ssler-Lorenz system} 

The system of equations (\ref{RLx1})-(\ref{RLy3}) is solved using a 4th order Runge-Kutta routine with time step $dt = 0.005$. The time series are generated by recording the variables every 6 time steps of integration and after an initial lapse of 500 steps, to avoid transients.
The embeddings we used in this case were
$(y_2(i+3) , y_2(i+2) , y_2(i+1) , x_2(i+2) , x_2(i+1) , x_2(i))$ for $TE_{x_2 \to y_2}$ and $(x_2(i+3) , x_2(i+2) , x_2(i+1) , y_2(i+2) , y_2(i+1) , y_2(i))$ for $TE_{y_2 \to x_2}$.

\subsubsection{Chain of coupled Lorenz systems}
%\end{enumerate}

The system of equations (\ref{eq1}-\ref{eq2}) is also solved using a 4th order Runge-Kutta method with the same integration step as for the R\"ossler-Lorenz system. The variables $x_1$, $x_2$ and $x_3$ are sampled using the same sampling time and initial lapse. To compute $TE_{x_i\to x_j}$, with $i\neq j\in \lc 1,2,3\rc$, we use the embedding $(x_j(t+3),x_j(t+2),x_j(t+1),x_i(t+2),x_i(t+1),x_i(t))$ while for the estimation of the conditional transfer entropy $TE_{x_1\to x_3|x_2}$ we used the embedding $(x_3(t+3),x_3(t+2),x_1(t+2),x_2(t+2),x_2(t+1),x_2(t))$. For the case 
$TE_{x_3\to x_1|x_2}$, the embedding is obtained as in the case of $TE_{x_1\to x_3|x_2}$ by simply interchanging the roles of $x_3$ and $x_1$.

\subsection{Bin sizes}

The choice of the size of the intervals along each axis is adapted to the size of the attractor and the number of points available. Following \cite{Krakovska2018}, if $N$ is the number of points furnishing the (embedded) attractor and $d$ is the embedding dimension, the number of intervals along each axis is taken as 
$N_{int} = {\rm min}\lc {\rm ceil}\lb N^{1/(d+1)}\rb , n_{max}\rc$, where ${\rm ceil(\cdot)}$ denotes the ceiling and $n_{max}$ is taken to be $9$ for $d=3$ and $4$ for $d=6$.
If $\lc p_i\rc$ is the set of points furnishing the attractor (it could either be the actual set of points in the embedding or the result of the sampling of the simplices in the triangulation), the size of the intervals along the $a$-th axis is chosen as follows: Let $O_a = \lb1-\frac{1}{10\,N_{int}}\rb {\rm min}\lc (p_i)_a | 1\leq i\leq N\rc $ and $T_a = \lb1+\frac{1}{10\,N_{int}}\rb{\rm max}\lc (p_i)_a | 1\leq i\leq N\rc$. $O$ will be referred to as the origin of the attractor. The interval size along the $a$-th axis is determined as $\epsilon_a = (T - O)_a/N_{int}$.  

The results for the KDE estimator are obtained using the minimum interval size for each case, that is ${\rm min}\lc \epsilon_a\rc$.

\subsection{Constructing the binning}

Suppose $E = \lc p_1,\cdots,p_N\rc\subset \mathbb R^d$ is the set of points furnishing the reconstructed attractor (in the case of the grid estimator) or the set of final sampling points (in the case of the triangulation estimator). Let $\epsilon = (\epsilon_a)$ and $O = (O_a)$ be the bin size and the origin of the attractor (see the previous section). Call $(x_1,\cdots,x_d)$ the coordinate axes on the embedding space, and generically denote as $A_{n+1}:=(x_1,\cdots,x_{n_1})$, $A_n:= (x_{n_1+1},\cdots,x_{n_2})$ and $B_n:=(x_{n_2+1},\cdots,x_{d})$, the variables on which the transfer entropy $TE_{B\to A}$ is computed. Each point $p_l\in E$ is assigned a unique triplet of integer tuples $I_l = (i^l_1,i^l_2,i^l_3)$, with $i^l_1 = (j^l_1,\cdots,j^l_{n_1})$, $i^l_2 = (k^l_{n_1+1},\cdots,k^l_{n_2})$ and $i^l_3 = (m^l_{n_2+1},\cdots,m^l_{d})$, and such that $O_a + (j^l_a-1)\epsilon_a < (p_l)_a\leq O_a + j^l_a\epsilon_a$, for all $1\leq a\leq n_1$ (analogously for $i^l_2$ and $i^l_3$). The unique elements in the set $\lc I_1,\cdots,I_N\rc$, say $\lc I_1,\cdots,I_M\rc$, identify the bins that contain at least one point from the set $E$ and constitute the binning used to compute the transfer operator and the TE (in the case of the grid estimator) and just the TE in the case of the triangulation estimator.

\subsection{Grouping of variables for TE computation}

For the case of the coupled logistic maps (both UCLM and BCLM), the TE corresponding to $x\to y$ uses the grouping of variables $A_{n+1} = (x(n+1))$, $A_n = (x(n))$ and $B_n = (y(n))$, while the TE corresponding to $y\to x$ is computed with the variable grouping $A_{n+1} = (y(n+1))$, $A_n = (y(n))$ and $B_n = (x(n))$.

For the case of the R\"ossler-Lorenz system, the gathering of variables used to compute $TE_{x_2\to y_2}$ is $A_{n+1} = (y_2(i+3))$,
$A_n = (y_2(i+2),y_2(i+1))$ and $B_n = (x_2(i+2),x_2(i+1),x_2(i))$. For computing $TE_{y_2\to x_2}$ we use the same gathering of variables but interchanging the symbols $x_2$ and $y_2$.

The TE corresponding to the coupling $x_i\to x_j$ (both direct and indirect) in section \ref{condTE} is computed using the gathering of variables
$A_{n+1} = (x_j(t+3))$, $A_n = (x_j(t+2))$ and $B_n = (x_i(t+2),x_i(t+1),x_i(t))$.

\subsection{Computation of conditional TE}

Given the variables $A_{n+1}$, $A_n$, $B_n$ and $C_n$, the conditional transfer entropy $TE_{B\to A|C}$ is computed as 

{\footnotesize 
\begin{equation}
    \int P(A_{n+1},A_n,B_n,C_n)\,\log\frac{P(A_{n+1}|A_n,B_n,C_n)}{P(A_{n+1}|A_n,C_n)}\,.\nonumber
\end{equation}
}

The conditional $TE_{x_1\to x_3|x_2}$ in section \ref{condTE} is computed using the gathering of variables $A_{n+1} = (x_3(t+3))$, $A_n = (x_3(t+2))$, $B_n = (x_1(t+2))$ and $C_n = (x_2(t+2),x_2(t+1),x_2(t))$. The conditional $TE_{x_3\to x_1|x_2}$ is computed using the same gathering of variables but interchanging $x_1$ and $x_3$.

\subsection{$k$ nearest neighbors counting}

For the case of kNN estimator of mutual information for dimension 3 or less, we used 5 nearest neighbors to compute $I(a,(b,c))$ while 10 nearest neighbors were used to compute $I(a,b)$, where $a,b$ and $c$ denote generic variables. For higher dimensions ($\geq 4$), the same number of nearest neighbors may be taken for both mutual informations (figure 16 in \cite{Kraskov2004}). 

\section{Ergodicity cross check}\label{app:ergodic}

As a way of testing the ergodicity of the invariant measure estimated with our method, we compare the temporal and spatial averages of several functions for the UCLM and for the R\" ossler-Lorenz system. In particular, we
consider the functions: $h_a:={\rm sech}\lb \sqrt{x^2+y^2}\rb$, $h_b:=\beta\lb 1+x^2,1+y^2\rb$ and $h_c=\psi\lb\sqrt{x^2+y^2}\rb$, for the 
coupled logistic maps, and $h_d:={\rm sech}\lb \sqrt{(x_2)^2+(y_2)^2}\rb$, $h_e:=\beta\lb 1+(x_2)^2,1+(y_2)^2\rb$ and $h_f:=\psi\lb\sqrt{(x_2)^2+(y_2)^2}\rb$, for the R\"ossler-Lorenz system, where $\beta(x,y)$ is the Euler $\beta$ function and $\psi(x)$ is the digamma function. There is no particular reason behind the choice of these functions, other than being complicated functions having no obvious connection with the systems.

\begin{figure}[!ht]
	\centering

    \includegraphics[width=0.5\textwidth]{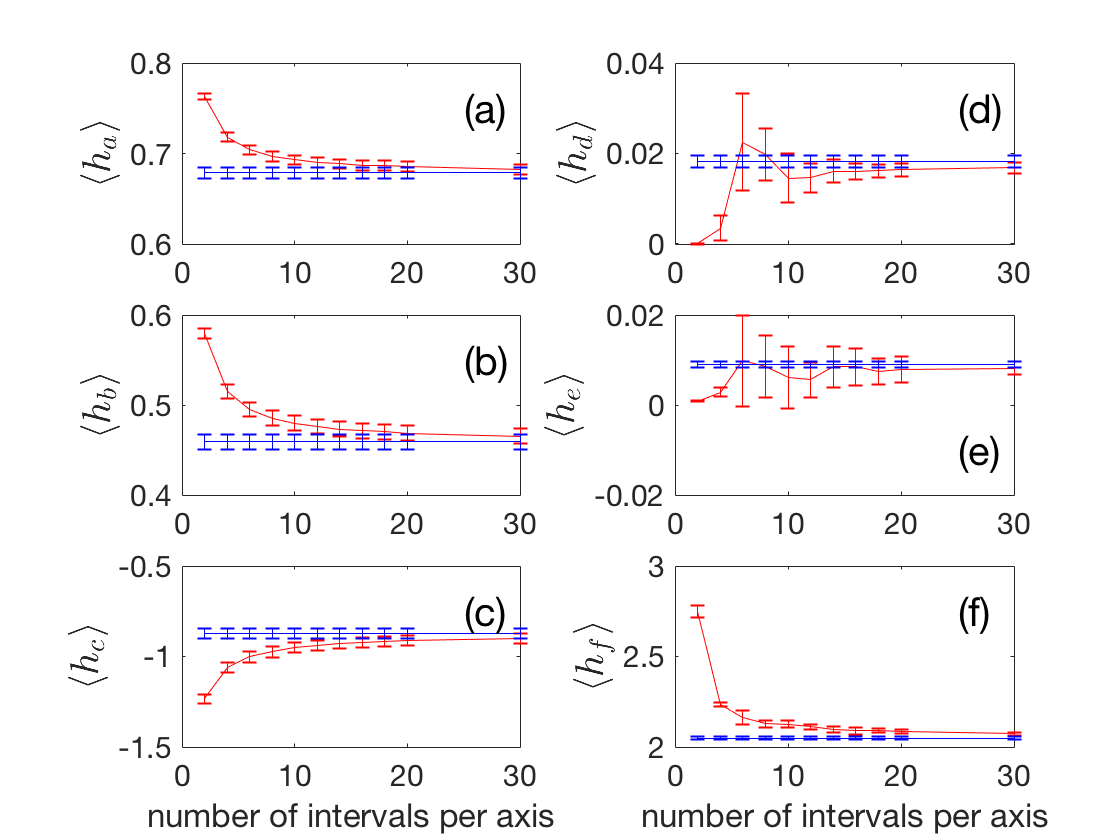}
	
	\caption{\footnotesize{Ergodicity test for 1000-point long orbits from the UCLM with $c=0.4$ and no noise ((a), (b) and (c)) and for 10000-point long orbits from the R\"ossler-Lorenz system with $c=1.5$ and no noise ((d), (e) and (f)). Blue lines show the temporal average and red lines the spatial average of the different functions (see text). The $x$-axis indicates the number of intervals that are taken along each axis in the embedding space for defining the grid. Error bars indicate the standard deviation over 50 realizations.}}
    \label{fig:ErgoCheck_CLM_RL}
\end{figure}

The spatial averages rapidly converge to the temporal averages as the bin size decreases (figure \ref{fig:ErgoCheck_CLM_RL}). Also, we point out that the rate of convergence seems to be fairly independent of the functions chosen to be averaged. This is maybe not so evident in the case of the R\"ossler-Lorenz system (figures \ref{fig:ErgoCheck_CLM_RL}d,e and f) but all the spatial averages seem to saturate beyond $\sim 15$ intervals per axis. This rate of convergence is rather dependent on the system and, likely more strongly, on the embedding dimension, suggesting that such a saturation could be used as a criterion for choosing a suitable bin size.

\section{Grid estimator vs visitation frequency estimator}\label{app:vf}

Here we compare the invariant density obtained using the grid estimator with the density that a direct visitation frequency estimation yields, as a function of the number of observations in the time series. For each instance of time series length, say $n$, we set a bin size (as explained in appendix \ref{app:binning}) and consider a partition into rectangular bins. We then apply the grid estimator to 50 realizations of time series with $n$ observations and generated from randomly chosen initial values. Hence, we obtain 50 estimates for the invariant distribution, say $\rho_{grid}(r,n)$, for $1\leq r\leq 50$. Using the same time series, we also compute the visitation frequency to each bin, obtaining thus $\rho_{vf}(r,n)$. We consider the discrepancy measure 
{\footnotesize
$$\delta(n) = \frac{1}{50}\sum_{r= 1}^{50}\frac{\parallel \rho_{grid}(r,n)-\rho_{vf}(r,n)\parallel}{{\rm max}\lc\parallel \rho_{grid}(r,n)\parallel\, ,\, \parallel \rho_{vf}(r,n)\parallel \rc}$$
}

with $\parallel v \parallel := {\rm max}\lc \abs{v_a}\rc$. 

\begin{figure}[!ht]
	\centering
    
    %\vspace{0.1cm}
    \includegraphics[width=0.5\textwidth]{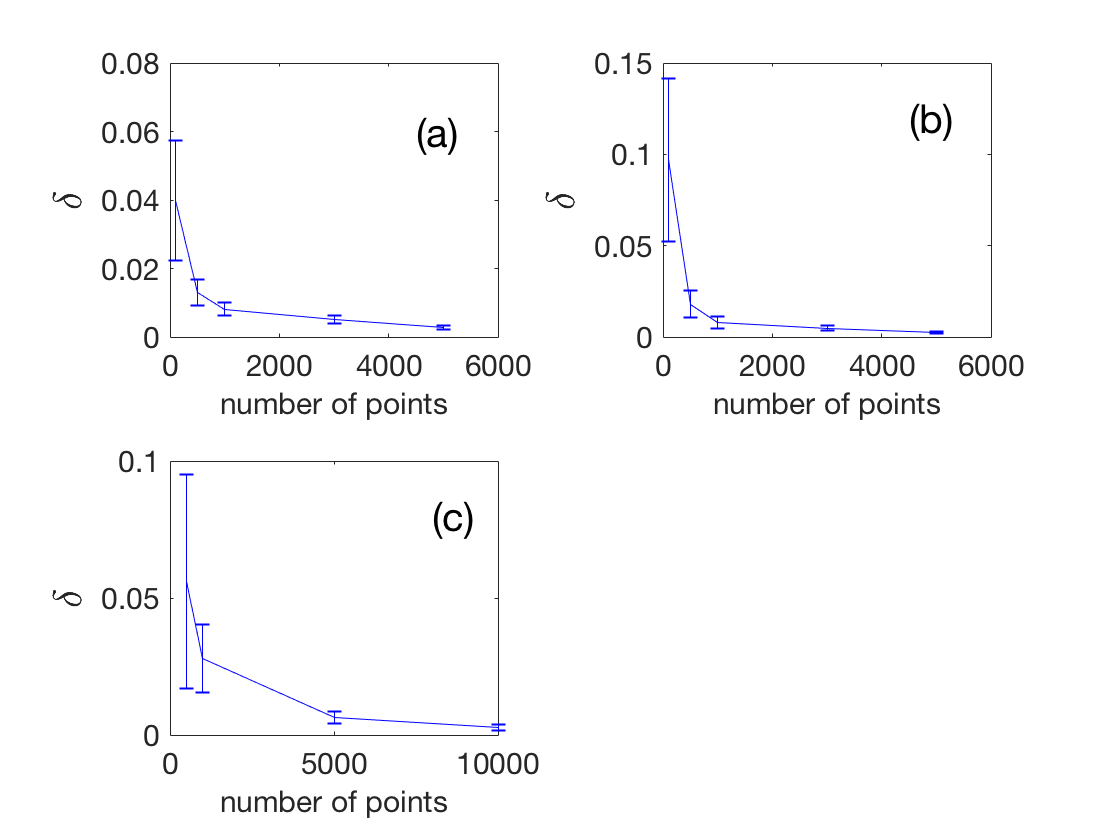}
    
    %\vspace{-1.5cm}
    
	\caption{\footnotesize{Mean value and standard deviation over 50 realizations of the discrepancy between the invariant densities computed with the grid estimator and via computation of the visitation frequency (appendix \ref{app:vf}), as a function of the number of observations in the time series for: (a) UCLM with $c = 0.4$ and no noise; (b) BCLM with $c=0.2$ and no noise; (c) R\" ossler-Lorenz system with $c=1.5$ and no noise.}}
    \label{fig:grid_vs_vf}
\end{figure}

We apply this procedure to our example systems and find that both methods for estimating invariant densities produce the same outcomes (within very small discrepancies) for long enough time series (figure \ref{fig:grid_vs_vf}). By virtue of the ergodic theorem, the invariant density of the transfer operator and the invariant density yielded by the frequency of visitations must coincide. 
Hence, the convergence of the density estimates is expected for long enough time series.

\section{Computing the intersecting volume between simplices in dimension $d$.}\label{app:PolytopeVol}

We outline the method used in this paper for computing the volume of the intersection between two simplices. Let $V = \lc p_0,\cdots,p_d\rc$ be $d+1$ affinely independent points in $\mathbb R^d$. The simplex $S$ with vertices $V$ is the convex hull of them, usually denoted as $S=CH(V)$, and defined as: all the points in $R^d$ constructed as $x = \alpha_0 p_0+\cdots+\alpha_d p_d$, with $\alpha_i\geq 0$ and $\alpha_0+\cdots+\alpha_d = 1$. In addition, $x$ lies in the interior of $S$, denoted as $x\in \mathring S$, if and only if all $\alpha_i > 0$. A generic boundary of $S$ is the simplex with vertices $\lc p_{\sigma_0} ,\cdots,p_{\sigma_k}\rc$, where $\lc\sigma_0 < \cdots < \sigma_k\rc$ is a (non-empty) selection of $\lc 0,\cdots,d\rc$, for $k = 0,\cdots,d$. The proper faces of the simplex correspond to $k=d-1$. Suppose $S_1 = CH(V_1)$ and $S_2 = CH(V_2)$, with $V_1 = \lc p_0,\cdots,p_d\rc$ and $V_2 = \lc q_0,\cdots,q_d\rc$. The method for computing the volume of the intersection $S_1\cap S_2$ used in this work is based on the following result (the proof of which is given at the end of this appendix):
\begin{theor}\label{theor1}
Let $S_1,S_2\subset \mathbb R^n$ be two simplices of dimensions $n\geq m\geq 1$, respectively, and with $S_1\cap S_2\neq\emptyset$. Let $\mathcal I$ be the set of points in $\mathbb R^n$ constructed as follows: $p\in\mathcal I$ if $p\in \mathring{B_1}\cap \mathring{B_2}$, where $B_1$ and $B_2$ are boundaries of $S_1$ and $S_2$, respectively, and not supporting any common direction. Then it holds that $S_1\cap S_2=CH(\mathcal I)$. 
\end{theor} 
In other words, if $\mathcal I = \lc x_1,\cdots,x_N\rc$ is such a set, then $S_1\cap S_2$ consists of all the points of the form 
$\beta_1 x_1+\cdots+\beta_N x_N$, for $\beta_n \geq 0$ and $\beta_1+\cdots+\beta_N = 1$.\\
The set $\mathcal I$ may be found as follows: let $B_1 = CH(\lc p_{\rho_1} ,\cdots,p_{\rho_r}\rc)$ and $B_2 = CH(\lc q_{\sigma_1} ,\cdots,q_{\sigma_s}\rc)$ and, with no loss of generality, assume that $r\geq s$. Reorder the vertices of $S_1$ as $ \lc p_{\rho_1} ,\cdots,p_{\rho_r},p_{\rho_{r+1}} ,\cdots,p_{\rho_{d+1}}\rc$. Every point in $\mathbb R^d$ can be expressed as a unique affine linear combination of these vertices, possibly with negative coefficients. In particular, $q_{\sigma_i} = \sum_{j=1}^{d-r+1}\gamma_{ji}\,p_{\rho_{r+j}} + \cdots$, where the coefficients on the rest of the vertices of $S_1$ are omitted. Denote the least dimensional affine space containing the boundary $B_1$ (respectively $B_2$) as $\Pi_1$ (respectively $\Pi_2$). 
By definition, the affine spaces generated by the sets of (affinely independent) points $\lc x_1,\cdots,x_n\rc $ and $\lc y_1,\cdots,y_m\rc$, intersect uniquely if the equations 
{\small \begin{align}
 & \sum_{i=1}^n \alpha_i x_i =\sum_{j=1}^m\beta_j y_j\,,\\
 & \sum_{i=1}^n \alpha_i  =\sum_{j=1}^m\beta_j = 1\,,
\end{align}}
\noindent have unique solution. 
In our case, this translates into the conditions
{\small \begin{align}
  &{\rm rank}\,\gamma = s-1\\
  &{\rm rank}\lb\begin{array}{c} \gamma\\ {\bf 1}\end{array}\rb = s
\end{align}}
where $\gamma$ is the $(d-r+1)\times s$ matrix with entries $\gamma_{ab}$ and ${\bf 1}$ denotes a row of ones. If $x$ is the unique intersecting point, then 
{\small \begin{equation}
x = \sum_{i=1}^r \alpha_i \,p_{\rho_i} = \sum_{j=1}^s \beta_j\,q_{\sigma_j}\,,
\end{equation}}
with $\sum_i \alpha_i = \sum_j \beta_j = 1$, and only if $\alpha_i,\beta_j> 0$, the point $x$ is in the interior of both boundaries and thus an element of $\mathcal I$.

Once the set $\mathcal I$ has been found it can be further decomposed as $\mathcal I = \cup_{a=1,2; j=0,\cdots, d}\,\mathcal I^{(a)}_j$ where $\mathcal I^{(a)}_j:=\mathcal I\cap F^{(a)}_j$ and $F^{(a)}_j$ denotes the face of the simplex $S_a$ that lies opposite to the $j$-th vertex of that simplex. Some of these subsets might be empty and there might be repetitions. Call $\mathcal F$ the set of the unique elements in the collection $\lc \mathcal I^{(a)}_j\rc_{a=1,2\,;\,j=0,\cdots,d}$ that appear at most once for $a=1$ or once for $a=2$. If the element $\mathcal I^{(1)}_i$ appears only for $a=1$ then it is easy to check that the intersection $F^{(1)}_i\cap S_2$ produces a convex set of dimension $d-1$, i.e. a proper face of $S_1\cap S_2$. On the other hand, theorem \ref{theor1} applied to the simplices $F^{(1)}_i$ and $S_2$ precisely yields the set of vertices $\mathcal I^{(1)}_i$. The same holds if the set of vertices $\mathcal I^{(2)}_i$ appears only for $a=2$. In the case where any such set, $\mathcal I^{(a)}_j$, appears for both $a=1$ and $a=2$, its convex hull is a proper face of $S_1\cap S_2$ only when the corresponding faces containing $\mathcal I^{(a)}_j$ are parallel. Each face of $S_1\cap S_2$ may be triangulated (some of them might be already simplices) and the triangulation of these faces, together with any point in the interior of $S_1\cap S_2$, for instance its centroid, form a triangulation of the intersection between the simplices. Its volume is then computed as the sum of the volumes of the simplices in this last triangulation.	

%\vspace{1cm}
\subsection{Proof of theorem \ref{theor1}}

The statement in theorem \ref{theor1} is actually a corollary of a well known result on convex geometry:  

\begin{theor}[{\bf Minkowski}]\label{theor2}
Every convex and compact set in $\mathbb R^n$ is the convex hull of its extreme points.
\end{theor}

%\vspace{.5cm}

The proof of theorem \ref{theor2} can be found in \cite{grunbaum2003convex}. A point $x$ in a convex set $P$ is said to be extreme if  the equality $x= \lambda\,y+(1-\lambda)\,z$, for $y,z\in P$ and $0<\lambda<1$, requires $x=y=z$. In other words, $x$ is not found in the interior of any segment contained in $P$. For our purposes, it is convenient to use an (easily shown to be) equivalent definition: $x$ is extreme if for every unit vector $u$ and for every $\epsilon>0$, there is $\abs{\lambda}<\epsilon$, such that 
$ x + \lambda \,u \notin P$. The set of extreme points of $P$ is denoted as $ext(P)$.
\begin{proof}[(Proof of theorem \ref{theor1}).]
 We claim that the set $\mathcal I$, as defined in theorem \ref{theor1}, equals $ext\lb S_1\cap S_2\rb$. To see this, let $\lc p_0,\cdots,p_n\rc$ be the vertices of $S_1$ and $\lc q_0,\cdots,q_m\rc$ be the vertices of $S_2$ and let $x\in \mathring{B_1}\cap\mathring{B_2}$ with $B_1$ and $B_2$, boundaries verifying the properties required in theorem \ref{theor1}. Next, let $u$ be an arbitrary unit vector and assume, without loss of generality (w.l.o.g.), that the direction $u$ is not supported by $\Pi_1$ (the least dimensional affine space supporting $B_1$). Further assume, w.l.o.g., that $\lc p_0,\cdots,p_{r}\rc$ are the vertices of the boundary $B_1$. Then $u = \sum_{i=1}^r \gamma_i (p_i-p_0) + \sum_{j=r+1}^n \mu_j (p_j-p_0)$ with not all $\mu_j$ vanishing. Assume, w.l.o.g., that $\mu_{r+1}\neq 0$. Therefore, $x+\lambda\,u = \lambda\,\mu_{r+1} \,p_{r+1} + \sum_{i=0}^r \alpha_i(\lambda)\, p_{i} + \sum_{j>r+1} \beta_j(\lambda)\, p_j$. Given $\epsilon>0$ arbitrary, take $\lambda = -{\rm sign}\lb \mu_{r+1}\rb\,\dfrac{\epsilon}{2}$, it then holds that $x+\lambda\,u = -\abs{\mu_{r+1}}\dfrac{\epsilon}{2} \,p_{r+1} +\cdots$ and therefore $x+\lambda\,u\notin S_1$. This shows that $x$ is an extreme point of $S_1\cap S_2$ and since $x\in \mathcal I$ was arbitrary, it follows that $\mathcal I\subset ext\lb S_1\cap S_2\rb$.  \\
 \\
To see the reverse inclusion, let $x\in ext\lb S_1\cap S_2\rb$ arbitrary. W.l.o.g., assume that $x = \sum_{i=0}^r \alpha_i p_i = \sum_{j=0}^s \beta_i q_i$ with all $\alpha_i >0$ and all $\beta_j >0$. Call $\Pi_r$ the least dimensional affine space containing the vertices $\lc p_0,\cdots,p_r\rc$ and $\Pi_s$ the affine space with the same property with respect to the vertices $\lc q_0,\cdots,q_s\rc$. It holds that $\Pi_r$ and $\Pi_s$ do not support any common direction. Indeed, suppose $u$ is a unit vector along a direction supported by both $\Pi_r$ and $\Pi_s$ then, having that all the coefficients $\alpha_i$ and $\beta_j$ are strictly positive, it follows that for some $\epsilon>0$ small enough, $x+\lambda\,u\in S_1\cap S_2$, for all $\abs{\lambda}<\epsilon$, contradicting that $x$ is an extreme point. This shows that $\mathcal I = ext\lb S_1\cap S_2\rb$. Given that both $S_1$ and $S_2$ are compact and convex sets, so it is $S_1\cap S_2$ and the proof is completed by using theorem \ref{theor2}. \\   
\\
\end{proof}

\section{Sample-and-filter approach.}\label{app:mixed}

\begin{figure}[!ht]
	\centering
    
    %\vspace{0.1cm}
    \includegraphics[width=0.5\textwidth]{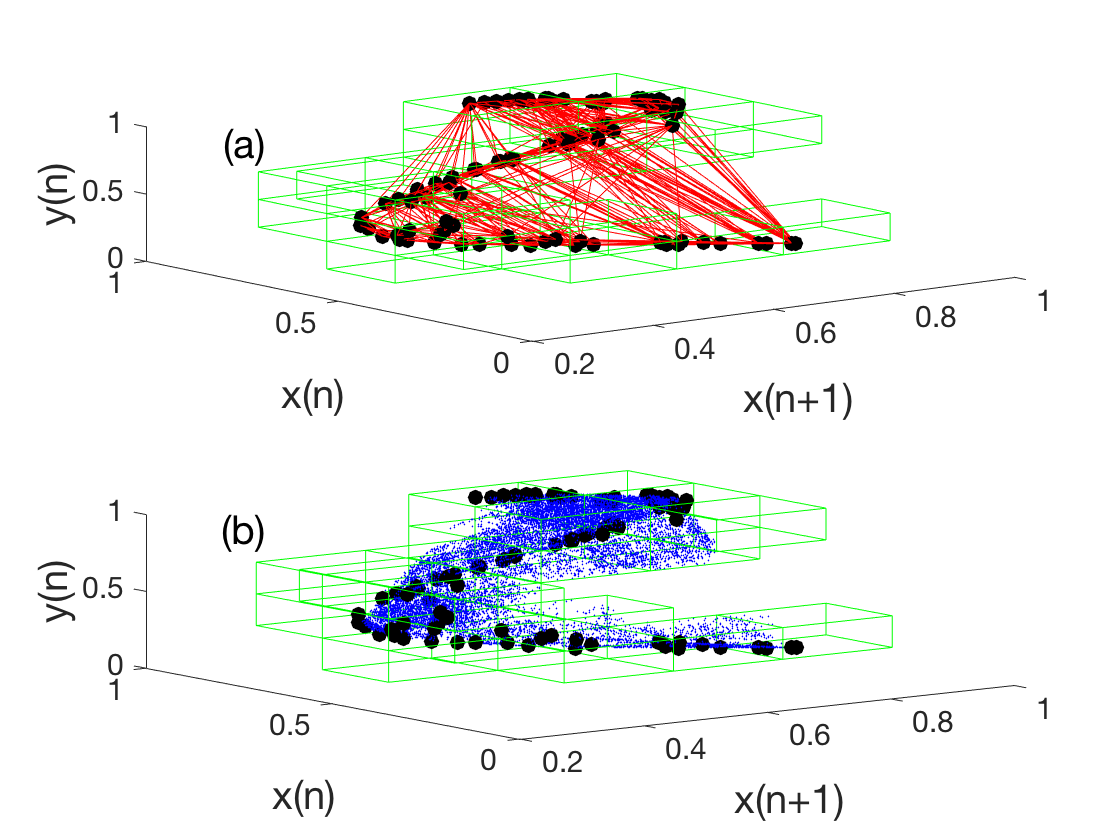}
    
    %\vspace{-1.5cm}
    
	\caption{\footnotesize{Example of the sampling and filtration of 100 embedding points for the UCLM system with $c=0.4$, no noise and using the embedding $(x(n+1) , x(n) , y(n))$. Solid black points represent the original embedding points and the grid is shown in green. (a) initial triangulation of the reconstructed attractor. (b) each simplex is sampled with 80 points (small blue points), where those not lying in the grid have been filtered out. The number of points after filtering (small blue points) is about 23000.}}
    \label{fig:filtering}
\end{figure}

Here we provide a tentative modification of the triangulation estimator in order to reduce its high computational demands. As described in \ref{sec:triang}, the triangulation estimator is based on an initial partition of the embedded attractor into simplices and the map generating the dynamics is approximated linearly onto each simplex \cite{Froyland1997}, say $\tilde\psi$. Given the initial triangulation of the attractor, we make use of the piecewise linear approximation of the map to generate sampling points as an input to the grid estimator (equation (\ref{grid_to})):

\begin{enumerate}
    \item Given an embedded attractor in a $d$ dimensional space (we assume it consists of few hundred points), let $\lc S_1,\cdots,S_N\rc$ be its triangulation into $d$-simplices and  $\lc B_1,\cdots,B_M\rc$ be the set of bins in a regular grid that are visited by the embedded points (figure \ref{fig:filtering}(a)). The size of the bins in the regular grid is adapted to the number of points in the reconstructed attractor (appendix \ref{app:binning}).   
    \item Each simplex is sampled with $N_s$ points using a predefined matrix of convex coefficients. More specifically, let $C$ be a $N_s\times (d+1)$ matrix such that $C_{a i} \geq 0$, no two rows are equal and $\sum_{j=1}^{d+1} C_{a j} = 1$, for all $1\leq a\leq N_s$ and $1\leq i\leq d+1$. Let $\lc v_1,\cdots,v_{d+1}\rc$ be the vertices of the simplex $S_n$ in the triangulation of the attractor and $\lc \tilde\psi(v_1),\cdots,\tilde\psi(v_{d+1})\rc$ be the vertices of the simplex being the image of $S_n$ under the map (see \ref{sec:triang} for details).
    The $a$-th sampling point of the simplex $S_n$ is given by
    $p_{n,a} := \sum_{j=1}^{d+1} C_{a j} v_j$ and its image under $\tilde\psi$ is given by $\tilde\psi(p_{n,a}) = \sum_{j=1}^{d+1} C_{a j} \tilde\psi(v_j)$ (the map $\tilde\psi$ is linear on each simplex). Call $P:=\lc p_{n,a}\rc_{1\leq n\leq N,\,1\leq a\leq N_s}$, the set of all the sampling points of the simplices in the triangulation and $\tilde \psi(P) =\lc \tilde\psi(p_{n,a})\rc$, the set of image points.
    \item From $P$, discard all those points not lying in $\cup_{i=1}^M B_i$. The resulting set, $\bar P$, 
    contains (possibly thousands of) points that are distributed more tightly to the volume occupied by the reconstructed attractor (figure \ref{fig:filtering}b).  
    \item Finally, a new bin size is adapted to the number of points in $\bar P$. Their images are found in the set $\tilde\psi(\bar P) = \left.\lc \tilde\psi(p_{n,a})\,\right|\, p_{n,a}\in \cup_i B_i\rc$. Therefore, the transfer operator may be approximated using equation (\ref{grid_to}), which using the above notation reads
    
    {\small \begin{equation}
P_{ij} \simeq \frac{\sharp\lc p_{n,a}\,|\, \tilde\psi(p_{n,a})\in B_j\,\cap\,p_{n,a}\in B_i\rc}{\sharp\lc p_{m,b}\,|\,p_{m,b}\in B_i\rc}\,.\nonumber
\end{equation}} 
    
\end{enumerate}

\begin{figure}[!ht]
	\centering
    
    %\vspace{0.1cm}
    \includegraphics[width=0.5\textwidth]{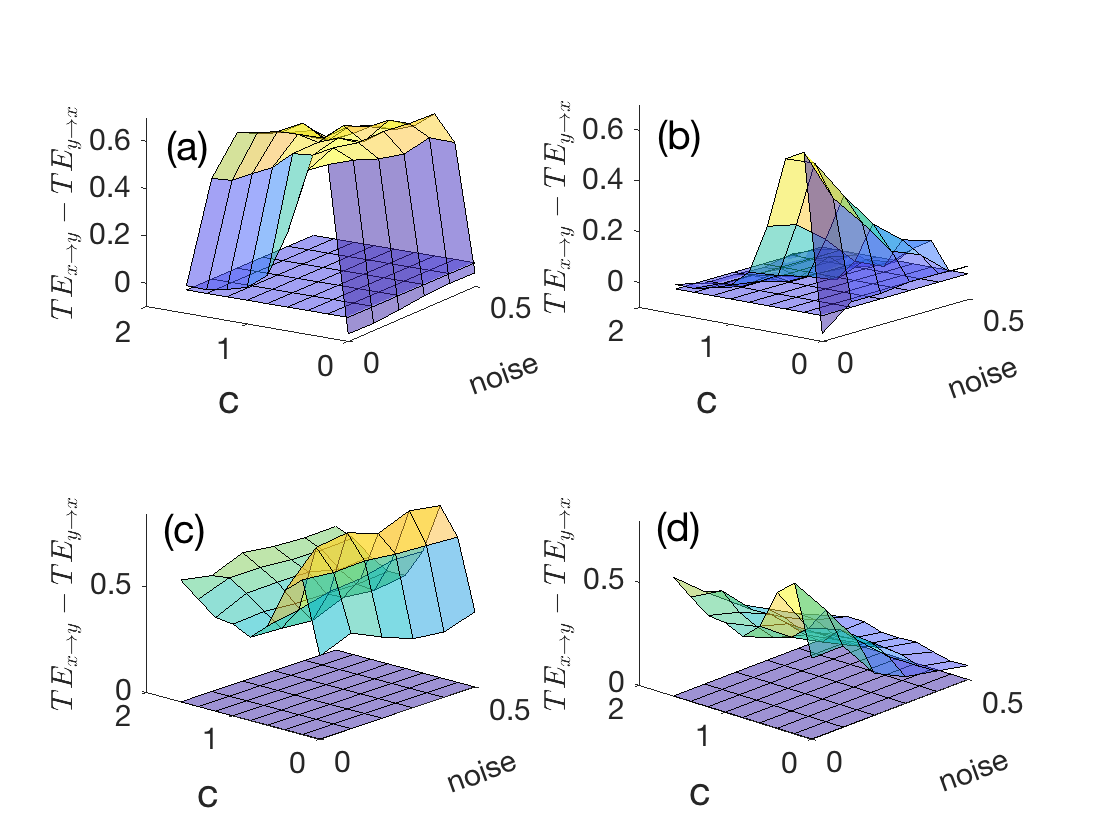}
    
    %\vspace{-1.5cm}
    
	\caption{\footnotesize{Mean value of $TE_{x\to y} - TE_{y\to x}$ over 50 realizations as a function of coupling constant and noise level obtained with the sample-and-filter approach. (a) UCLM with dynamical noise; (b) UCLM with measurement noise; (c) BCLM with dynamical noise; (d) BCLM with measurement noise.}}
    \label{fig:CLM_mixed}
\end{figure}

\begin{figure}[!ht]
	\centering
    
    %\vspace{0.1cm}
    \includegraphics[width=0.5\textwidth]{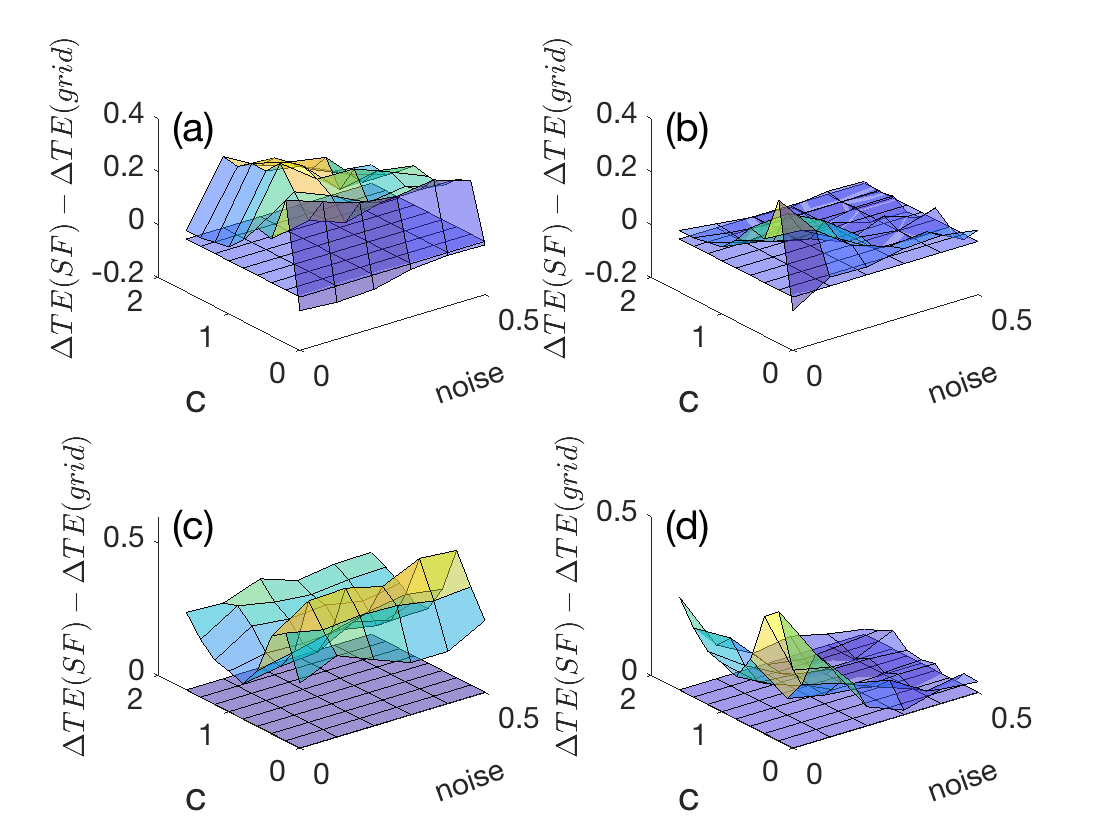}
    
    %\vspace{-1.5cm}
    
	\caption{\footnotesize{$\Delta TE(SF)- \Delta TE(grid)$ ($SF$ indicates sample-and-filter method) as a function of the coupling constant and noise level for: (a) UCLM with dynamical noise; (b) UCLM with measurement noise; (c) BCLM with dynamical noise; (d) BCLM with measurement noise.}}
    \label{fig:CLM_mixed_grid}
\end{figure}

\begin{figure}[!ht]
	\centering
    
    %\vspace{0.1cm}
    \includegraphics[width=0.5\textwidth]{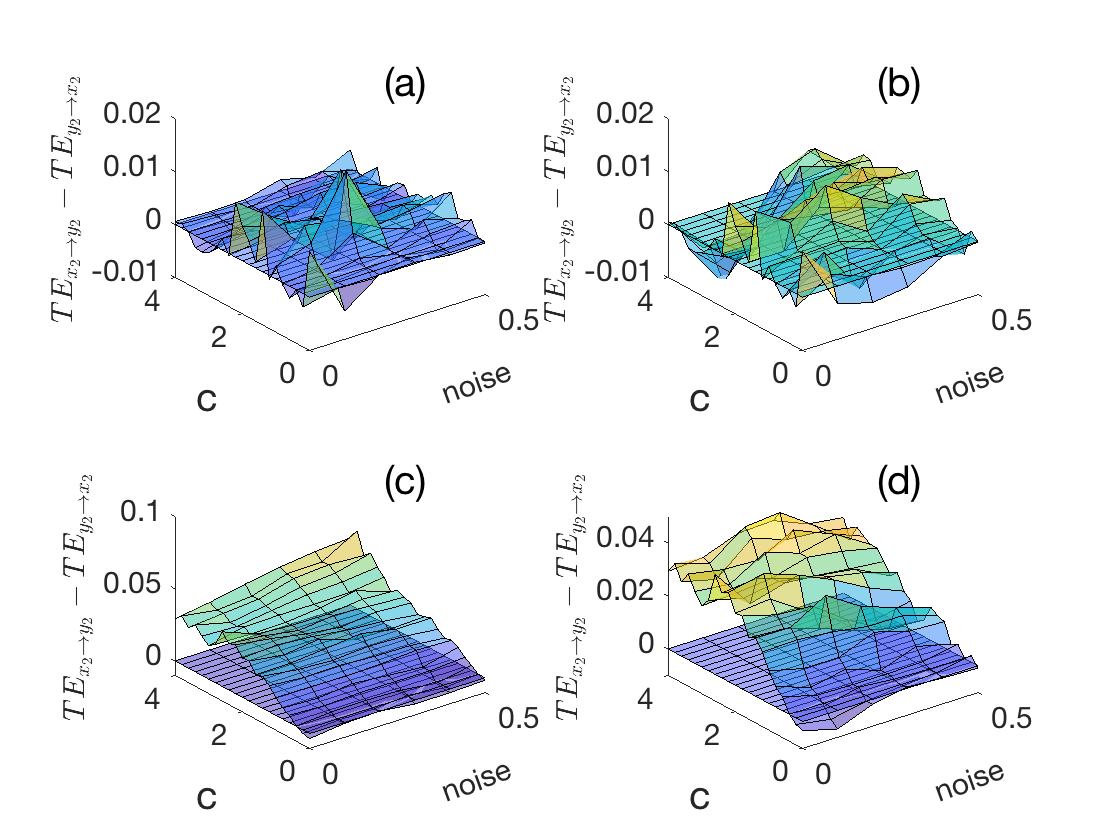}
    
    %\vspace{-1.5cm}
    
	\caption{\footnotesize{Mean value of $ TE_{x_2\to y_2}- TE_{y_2\to x_2}$ over 50 realizations as a function of the coupling constant and noise level for the R\"ossler-Lorenz system and: (a)  with dynamical noise obtained with the sample-and-filter approach; (b) the same as in (a) but with measurement noise instead; (c) obtained with the grid estimator and adding dynamical noise; (d) the same as in (c) but with measurement noise instead.}}
    \label{fig:RL_mixed_grid}
\end{figure}

 Note that generating sampling points in this manner only assumes piecewise linearity of the map and does not introduce any further bias. 
 We applied this approach to both UCLM and BCLM, using time series with 100 observations and 50 realizations from randomly chosen initial conditions. The dependence of $\Delta TE := TE_{x\to y} - TE_{y\to x}$
on the coupling constant and noise
is comparable to that obtained from the grid estimator (figures \ref{fig:CLM_mixed}a,b). Interestingly, the values for the $\Delta TE$ obtained with the sample-and-filter approach are, in general, higher than those obtained with the grid estimator (figure \ref{fig:CLM_mixed_grid}). This result suggests that from a sparse time series one may generate thousands of points from a piecewise linear approximation and obtain reliable results for the TE. We also apply the sample-and-filter approach to the R\" ossler-Lorenz system, in which case the results are less impressive (figure \ref{fig:RL_mixed_grid}). Note however, that these analyses used time series with only 200 observations in embedding dimension 6 (appendix \ref{app:binning}).            

%\newpage

%\newpage

%\newpage

%\section{Acknowledgements}

%\newpage

\section{Computational times for the triangulation estimator}\label{app:timing}

\begin{figure}[!ht]
	\centering
    
    %\vspace{0.1cm}
    \includegraphics[width=0.5\textwidth]{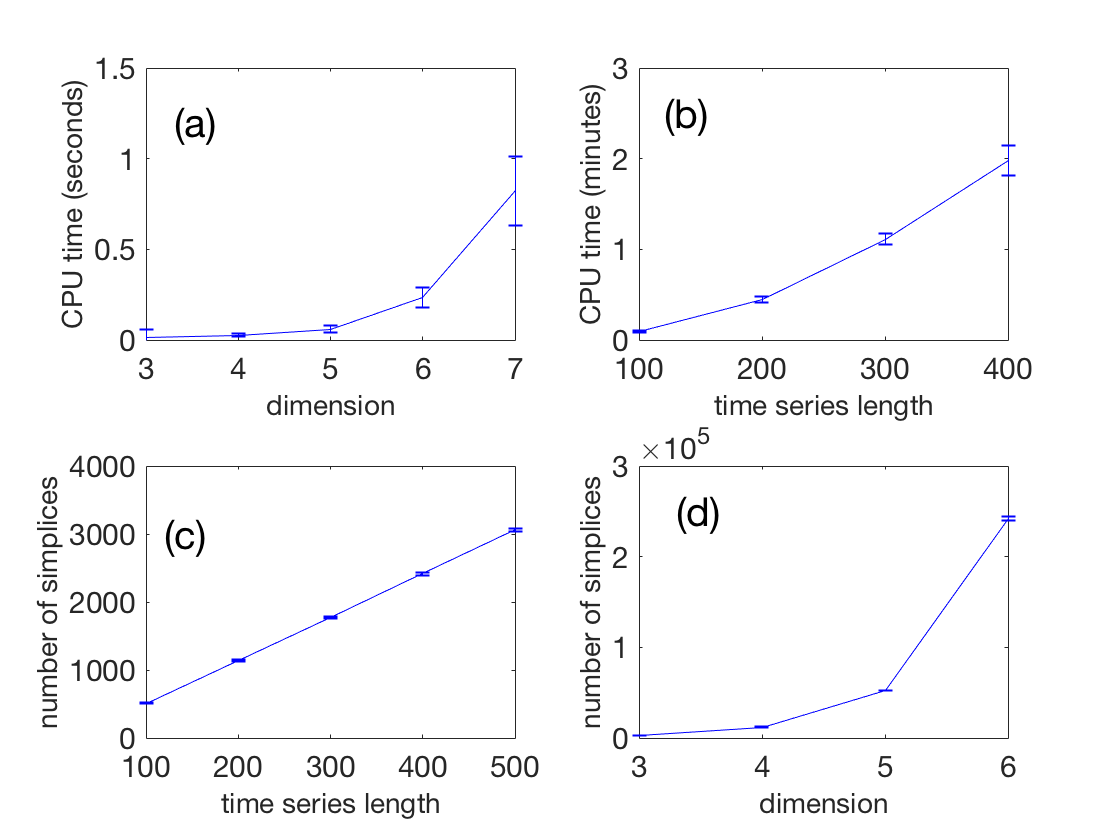}
    
    %\vspace{-1.5cm}
    
	\caption{\footnotesize{(a): mean and standard deviation over 50 realizations of the CPU time needed to obtain the volume of a non-trivial simplex intersection; (b) mean and standard deviation over 10 realizations of the CPU time required to obtain an estimate of the transfer operator for to the UCLM system with $c=0.4$ and no noise, in embedding dimension 3. The timing values were obtained using the {\it cputime} function in Matlab; (c) mean and standard deviation over 50 realizations of the number of simplices in a triangulation in 3d; (d) the same as in (c), fixing the number of observations in the time series to 500.}}
    \label{fig:timing}
\end{figure}

In this appendix we provide CPU times required to obtain the transfer operator using the triangulation estimator (figure \ref{fig:timing}b) as well as the CPU time required for obtaining the volume of a (non-trivial) simplex intersection (figure \ref{fig:timing}a). 
%The transfer operator is numerically computed as a sparse matrix, using a parallel loop in Matlab with 28 workers, where the parallelization is done over the simplices in the triangulation. 
The simulations were run in Matlab using a MacBook Pro with a 2.8 GHz Intel Core i7 processor. Computation times increase for higher dimension because the CPU time to obtain the volume for a simplex intersection and the number of simplices in the triangulation both scale exponentially (figures \ref{fig:timing}a,d). In particular, for dimension 5, the CPU time required to obtain the volume of a non-trivial simplex intersection is $t_s \sim 5\cdot 10^{-2} \,s$ and the typical number of simplices in a triangulation generated from 500 points is $n_s\sim 5\cdot 10^4$. If each simplex intersects non-trivially with just $10$ of the simplices in the triangulation (which is a quite optimistic estimate) we are left with a computation time for obtaining the transfer operator in the order
$t_{TO} \sim 10\,n_s\,t_s \sim 2.5\cdot 10^4 s \sim 7\,h$. For comparison, our grid estimator applied to 500-point long time series in dimension 5 requires a CPU time of $\sim 5\cdot 10^{-3}\,s$ to estimate the transfer operator.  

\newpage

%\bibliography{References.bib}


\end{document}